\documentclass[11pt,a4paper]{article}

\usepackage{jheppub}
\usepackage{latexsym}
\usepackage{multirow}
\usepackage{color}
\usepackage[usenames,dvipsnames,table]{xcolor}
\usepackage{graphicx}
\usepackage{epsfig}  
\usepackage{epsf}    
\usepackage{dcolumn}
\usepackage{bm}
\usepackage{dcolumn}
\usepackage{textcomp}
\usepackage{float}
\usepackage{subfig}
\usepackage{relsize}
\usepackage{multirow}
\usepackage[toc,page]{appendix}

\usepackage{lineno}

\usepackage{hypcap}
\usepackage[]{hyperref}
\usepackage{makecell}
\usepackage{color}
\usepackage{pifont}
\usepackage{appendix}
\usepackage{amsmath}
\usepackage{multirow,bigdelim}  
\usepackage{lineno}
\usepackage[normalem]{ulem}
\hypersetup{
  bookmarks=true,         
  unicode=false,          
  pdftoolbar=true,        
 pdfmenubar=true,        
 pdffitwindow=true,     
 pdfstartview={FitH},    
 pdfsubject={Neutrino Oscillations Phenomenology},   
 pdfnewwindow=true,      
 pdfcreator={RevTeX},
 colorlinks=true,       
 linkcolor=red,          
 citecolor=blue,        
 filecolor=black,      
 urlcolor=blue,           
  }

\newcommand{\be}{\begin{equation}}
\newcommand{\ee}{\end{equation}}
\newcommand{\ba}{\begin{eqnarray}}
\newcommand{\ea}{\end{eqnarray}}
\newcommand{\ldm}{\ensuremath{{\Delta m_{31}^2}}}         
\newcommand{\sdm}{\ensuremath{{\Delta m_{21}^2}}}
\newcommand{\eps}{\varepsilon}

\newcommand{\ms}{\Delta m^2_{21}}
\newcommand{\ma}{\Delta m^2_{31}}

\newcommand{\eee}{\varepsilon_{ee}}
\newcommand{\eem}{\varepsilon_{e\mu}}
\newcommand{\eems}{\varepsilon^\ast_{e\mu}}
\newcommand{\eet}{\varepsilon_{e\tau}}
\newcommand{\eets}{\varepsilon^\ast_{e\tau}}
\newcommand{\emm}{\varepsilon_{\mu\mu}}
\newcommand{\emt}{\varepsilon_{\mu\tau}}
\newcommand{\emts}{\varepsilon^\ast_{\mu\tau}}
\newcommand{\ett}{\varepsilon_{\tau\tau}}

\newcommand{\capdef}{}
\newcommand{\mycaption}[2][\capdef]{\renewcommand{\capdef}{#2}
       \caption[#1]{{\footnotesize #2}}}
\makeatletter
\renewcommand{\fnum@table}{\textbf{\tablename~\thetable}}
\renewcommand{\fnum@figure}{\textbf{\figurename~\thefigure}}
\makeatother

\preprint{IP/BBSR/2025-01}

\title{Probing Earth's core using atmospheric neutrino oscillations in the presence of NSI at INO-ICAL}

\author[a,b]{Krishnamoorthi J,}
\author[a,b]{Anuj Kumar Upadhyay,}
\author[a,c,d]{Anil Kumar,}
\author[a,d,e]{Sanjib Kumar Agarwalla}

\affiliation[a]{Institute of Physics, Sachivalaya Marg, Sainik School Post,
	Bhubaneswar 751005, India}
\affiliation[b]{Department of Physics, Aligarh Muslim University, Aligarh-202002, India }
\affiliation[c]{Applied Nuclear Physics Division, Saha Institute of
  Nuclear Physics, Block AF, Sector 1, Bidhannagar, Kolkata 700064, India}
\affiliation[d]{Homi Bhabha National Institute, Anushakti Nagar,
  Mumbai 400094, India}
\affiliation[e]{Department of Physics \& Wisconsin IceCube Particle Astrophysics Center, University of Wisconsin, Madison, WI 53706, U.S.A}

\emailAdd{krishnamoorthi.j@iopb.res.in  (ORCID: 0009-0006-1352-2248)}
\emailAdd{anuju@iopb.res.in  (ORCID: 0000-0003-1957-2626)}
\emailAdd{anil.k@iopb.res.in (ORCID: 0000-0002-8367-8401)}
\emailAdd{sanjib@iopb.res.in (ORCID: 0000-0002-9714-8866)}

\abstract{
Neutrinos can serve as a complementary and independent tool to gravitational and seismic studies in exploring the interior of Earth, thanks to their unique properties: extremely low interaction cross sections and flavor oscillations. With the precise measurements of neutrino oscillation parameters and observation of the non-zero value of mixing angle $\theta_{13}$, it has become feasible to detect the forward scattering of GeV-energy atmospheric neutrinos passing through Earth with ambient electrons in the form of matter effects on neutrino oscillation probabilities. These matter effects depend on both the neutrino energy and electron density distribution along their path, making them ideally suited for exploring the inner structure of Earth. Furthermore, in the presence of non-standard interactions (NSI) of neutrinos with matter, oscillation patterns undergo additional modifications. In this study, we quantify the capability of an atmospheric neutrino experiment, such as a magnetized iron calorimeter detector, to validate the Earth's core and measure the position of the core-mantle boundary in the presence of NSI. We perform this study considering a three-layered density profile of Earth. Our analysis demonstrates that neutrino non-standard interactions impact these Earth tomography measurements in comparison to standard interactions.	  
}

\arxivnumber{2507.02167}

\begin{document}
\maketitle
\flushbottom

\section{Introduction and motivation}
\label{sec:introduction}

The discovery of the neutrino oscillation phenomenon at the Super-Kamiokande experiment in 1998~\cite{SK:1998} proves the existence of the non-zero neutrino masses and mixing among different neutrino flavors, which is the first experimental evidence for theories beyond the standard model (BSM) of particle physics. The standard three-flavor neutrino oscillation phenomenon is described by six fundamental parameters~\cite{ParticleDataGroup:2020ssz}: a) the three mixing angles: solar mixing angle $\theta_{12}$, reactor mixing angle $\theta_{13}$, and atmospheric mixing angle $\theta_{23}$, b) two independent mass-squared differences: solar mass-splitting $\Delta m^2_{21}$ ($\equiv m^2_2 - m^2_1$) and atmospheric mass-splitting $\Delta m^2_{31}$ ($\equiv m^2_3 - m^2_1$ ), and c) one Dirac CP phase $\delta_{\text{CP}}$. Using the data collected by various neutrino experiments over the last twenty years, most of the oscillation parameters have been measured precisely~\cite{Capozzi:2025wyn,Esteban:2024eli,NuFIT,Capozzi:2021fjo,deSalas:2020pgw}, except the three unknowns: (i) the octant of $\theta_{23}$, (ii) the value of $\delta_\text{CP}$, and (iii) the sign of $\Delta m^2_{31}$ or neutrino mass ordering. The precise measurement of oscillation parameters, especially the discovery of the non-zero value of the reactor mixing angle $\theta_{13}$ by the Daya Bay experiment in 2012~\cite{DayaBay:2012fng}, has provided an opportunity to measure the matter effects encountered by the upward-going atmospheric neutrinos passing through Earth. The measurement of matter effects is expected to play an important role in several pressing issues, such as measuring the neutrino mass ordering, revealing the octant of $\theta_{23}$, exploring interesting BSM scenarios, and unraveling the information about the inner structure of Earth independent and complementary to gravitation and seismic studies.

Several BSM models have been developed to accommodate non-zero masses and mixing of neutrinos. The extension of the Standard Model (SM), in addition to the standard interactions (SI), allows subdominant interactions of neutrinos with matter fields that are not described by the SM. Such additional interactions are named as the non-standard interactions (NSI) of neutrinos. If NSIs exist in nature, it is very interesting from the phenomenological point of view to see their impact on neutrino production, propagation, and detection in a given experiment. The consequence of NSI on three-flavor neutrino oscillations are remarkable and extensively studied in the literature~\cite{Wolfenstein:1977ue,VALLE:1987,GUZZO:1991,Guzzo:2001,HUBER:2001,Fornengo:2001,Guzzo:2002,Gonzalez-Garcia:2004,Kopp:2008,Kopp:2010,Mitsuka:2011,Gonzalez-Garcia:2011,Ohlsson:2013,Gonzalez-Garcia:2013,Esmaili:2013,Choubey:2015,Miranda:2015,Mocioiu:2015,Salvado:2016,Farzan:2018,IceCube:2018,Esteban:2018,Proceedings:2019,Khatun:2020,Kumar:2021,HernandezRey:2021,Agarwalla:2021,IceCube:2021,ANTARES:2022}. The non-standard interactions of neutrinos can be classified into two categories: charged-current (CC) NSI and neutral-current (NC) NSI. The CC-NSI affects the neutrino fluxes at the production stage and neutrino interaction cross section at the detection level, while the NC-NSI can modify the neutrino propagation through matter.

As atmospheric neutrinos travel through Earth, they experience the standard matter effects due to the forward scattering with the ambient electrons. These matter effects depend upon the energy of neutrinos and the electron density distribution they encounter during their journey through Earth. While passing through the mantle, the neutrinos with energies in the range of 6 to 10 GeV experience the Mikheyev-Smirnov-Wolfenstein (MSW) resonance~\cite{Wolfenstein:1977ue,Mikheev:1986gs,Mikheev:1986wj}. On the other hand, the core-passing neutrinos with energies in the range of 3 to 6 GeV encounter additional matter effect which is known as parametric resonance (PR)~\cite{Ermilova:1986,Akhmedov:1988kd,Krastev:1989,Akhmedov:1998ui,Akhmedov:1998xq} or neutrino oscillation length resonance (NOLR)~\cite{Petcov:1998su,Chizhov:1998ug,Petcov:1998sg,Chizhov:1999az,Chizhov:1999he}. These density-dependent matter effects can alter the neutrino oscillation probabilities, and hence, atmospheric neutrinos can be an appealing tool for getting information about the internal structure of Earth. The provided information would be independent and complementary to that obtained from indirect methods, such as the gravitational measurements~\cite{Ries:1992,Luzum:2011,Rosi:2014kva,astro_almanac,Williams:1994,Chen:2014} and seismic studies~\cite{Gutenberg:1914,Robertson:1966,Volgyesi:1982,Loper:1995,Alfe:2007,Stacey_Davis:2008,McDonough:2022,Thorne:2022,Hirose:2022,McDonough_MMTE:2023}. 

The study of seismic wave propagation through different regions inside Earth indicates a layered structure in the form of concentric spherical shells. The inner structure of Earth can be broadly classified into two major layers: the mantle and the core, each containing multiple sub-layers. In the present study, we consider two sub-layers of the mantle, the outer and inner mantle, while the core is taken as a single layer. The densities of layers increase as we move deeper inside towards the center of Earth. However, this increase in density is not always smooth; sharp-density transitions occur at the boundaries between two adjacent layers. The most prominent density transition is observed between the inner mantle and the core. This transition region is known as the core-mantle boundary (CMB), which is located around $R_\text{CMB} = 3480 \pm 5$ km~\cite{Gutenberg:1914,young1987core,Masters1995,McDonough:2003,McDonough:2017,McDonough:2024}. The seismic studies also provide information about the density distribution inside the Earth. The most widely-used Earth density model, the Preliminary Reference Earth Model (PREM)~\cite{Dziewonski:1981xy}, has been developed using the seismic wave propagation data. The density of any layer in the PREM model is given as a one-dimensional function of radial distance of the layer from the centre of the Earth. In our study, we consider the density within a layer to be uniform, as the neutrino oscillations are not expected to be sensitive to small variations in density within a layer.

Although extensive data from seismic studies and gravitational measurements have significantly enhanced our understanding of the interior of Earth, there are still many open issues that remain unanswered. 
For instance, the uncertainties regarding the mass and the chemical composition of the core, as well as the density jump at the boundary between the outer core and the inner core, are still present~\cite{McDonough:2024}. Additionally, the precise amount of light elements, such as hydrogen, present inside the core remains uncertain~\cite{Williams:2001,Hirose:2021,Hirose:2022}. Moreover, the complementary and independent searches using probes such as geoneutrino detection~\cite{Araki:2005qa,Fiorentini:2007te,Bellini:2013wsa,McDonough2015,Michael2017,Bellini:2021sow}, neutrino absorption~\cite{Winter:2006vg,Gonzalez-Garcia:2007wfs,Donini:2018tsg}, and neutrino oscillations~\cite{Winter:2015zwx} could further improve our knowledge about the inner structure of Earth. The original idea of exploiting the absorption of high energy neutrino flux with energy more than a few TeV~\cite{Gandhi:1995tf,IceCube:2017roe} to explore the interior of Earth has been discussed in refs.~\cite{Placci:1973,Volkova:1974xa}. Later, the detailed studies using the attenuation of neutrinos from various sources, such as man-made neutrinos~\cite{Placci:1973,Volkova:1974xa,Nedyalkov:1981,Nedyalkov:1981pp,Nedyalkov:1981yy,Nedialkov:1983,Krastev:1983,DeRujula:1983ya,Wilson:1983an,Askarian:1984xrv,Volkova:1985zc, Tsarev:1985yub,Borisov:1986sm,Tsarev:1986xg,Borisov:1989kh,Winter:2006vg}, extraterrestrial neutrinos~\cite{Wilson:1983an,Kuo:1995,Crawford:1995,Jain:1999kp,Reynoso:2004dt,Francener:2024bfm}, and atmospheric neutrinos~\cite{Gonzalez-Garcia:2007wfs,Borriello:2009ad,Takeuchi:2010,Romero:2011zzb,Donini:2018tsg,Francener:2024bfm}, have been performed. Further, the studies using the matter effects in neutrino oscillations to learn about the Earth's interior have also been carried out using man-made neutrino beams~\cite{Ermilova:1986ph,Nicolaidis:1987fe,Ermilova:1988pw,Nicolaidis:1990jm,Ohlsson:2001ck,Ohlsson:2001fy,Winter:2005we,Minakata:2006am,Gandhi:2006gu,Tang:2011wn,Arguelles:2012nw}, supernova neutrinos~\cite{Lindner:2002wm,Akhmedov:2005yt,Hajjar:2023knk}, solar neutrinos~\cite{Ioannisian:2002yj,Ioannisian:2004jk,Akhmedov:2005yt,Ioannisian:2015qwa,Ioannisian:2017chl,Ioannisian:2017dkx,Bakhti:2020tcj}, and atmospheric neutrinos~\cite{Agarwalla:2012uj,IceCube-PINGU:2014okk,Rott:2015kwa,Winter:2015zwx,Bourret:2017tkw,Bourret:2019wme,Bourret:2020zwg,DOlivo:2020ssf,Kumar:2021faw,Maderer:2021aeb,Denton:2021rgt,Kelly:2021jfs,Capozzi:2021hkl,DOlivoSaez:2022vdl,Maderer:2022toi,Upadhyay:2022jfd,Upadhyay:2021kzf,Upadhyay:2024gra,Jesus-Valls:2024tgd,DOlivo:2025iqo}.

By exploiting the Earth's matter effects in atmospheric neutrino oscillations, several sensitivity studies have been performed recently in the context of current and future atmospheric neutrino experiments such as the IceCube~\cite{IceCube:2016zyt}, DeepCore~\cite{IceCube:2011ucd, IceCube:2016zyt}, Oscillation Research with Cosmics in the Abyss (ORCA)~\cite{KM3Net:2016zxf}, Precision IceCube Next Generation Upgrade (PINGU)~\cite{IceCube-PINGU:2014okk}, Hyper-Kamiokande (Hyper-K)~\cite{Hyper-Kamiokande:2018ofw}, Deep Underground Neutrino Experiment (DUNE)~\cite{DUNE:2021tad}, and Iron Calorimeter (ICAL)~\cite{ICAL:2015stm}. These studies encompass validating the CMB using ICAL~\cite{Kumar:2021faw,Anil_MMTE:2023}, determining the location of the CMB with DUNE~\cite{Denton:2021rgt} and ICAL~\cite{Upadhyay:2022jfd,Anil_MMTE:2023,Anuj_MMTE:2023}, constraining the density jump and core radius simultaneously using ICAL~\cite{Upadhyay:2024gra}, estimating the average densities of the core and mantle with ORCA~\cite{Winter:2015zwx,Maderer:2021aeb,Capozzi:2021hkl}, DUNE~\cite{Kelly:2021jfs}, ICAL~\cite{Raikwal:2023jkf}, and Hyper-K~\cite{Jesus-Valls:2024tgd}, probing the potential presence of dark matter within Earth using ICAL~\cite{Upadhyay:2021kzf,Anil_MMTE:2023}, and investigating the chemical composition of Earth's core using PINGU~\cite{IceCube-PINGU:2014okk}, Hyper-K and IceCube~\cite{Rott:2015kwa}, and ORCA~\cite{Bourret:2017tkw,Bourret:2019wme,Bourret:2020zwg,Maderer:2021aeb,Maderer:2022toi,DOlivoSaez:2022vdl,DOlivo:2025iqo}. The potential of the IceCube DeepCore detector to establish the Earth's matter effect, validate the non-homogeneous density profile inside Earth, and measure the mass of Earth and the correlated densities of different layers inside Earth using oscillations of weakly interacting neutrino at GeV energies is presented in refs.~\cite{Anuj_PPC:2024,Krishnamoorthi_PPC:2024,Sharmistha_PPC:2024,Chattopadhyay:2025cgt}.

We perform this study using the expected atmospheric neutrino events at a 50 kt magnetized Iron Calorimeter (ICAL) detector at the India-based Neutrino Observatory (INO)~\cite{ICAL:2015stm}. ICAL is optimized to detect atmospheric neutrinos and antineutrinos in the multi-GeV energy range, where the effects of the MSW and PR/NOLR resonances on neutrino oscillations are prominent. Due to its good angular resolution, it would be able to detect core-passing and mantle-passing neutrinos. The good energy resolution would help the ICAL detector to efficiently observe the MSW and PR/NOLR resonances. Thanks to the magnetic field, ICAL would be able to distinguish between neutrinos and antineutrinos by detecting $\mu^-$ and $\mu^+$ events separately and, hence, would be able to enhance the sensitivity to the matter effects. Since these matter effects depend upon the amount of density jump at the CMB and its location, ICAL would be sensitive to both of these features. In ref.~\cite{Kumar:2021faw}, some of the present authors have shown the potential of the ICAL detector to validate a high-density core inside Earth with a given density jump at the standard location of CMB. As a follow-up work in ref.~\cite{Upadhyay:2022jfd}, some of the present authors quantified the sensitivity of the ICAL detector to determine the location of CMB, considering various scenarios.

In the present study, we quantify the expected sensitivity of the ICAL detector to determine the presence of a high-density core and measure the location of the CMB in the presence of NSI. We investigate the effects of the flavor-violating NC-NSI parameters $\varepsilon_{e\mu}$, $\varepsilon_{e\tau}$, and $\varepsilon_{\mu\tau}$ on these measurements one at a time, i.e., we only allow one NSI parameter to be non-zero in our analysis, and the other parameters remain zero. We consider these parameters to be real, allowing both positive and negative values.

We organize this paper in the following way. In section~\ref{sec:formalism_nsi}, we discuss the theoretical formalism of NSI. In section~\ref{sec:earth-profile}, we demonstrate Earth density profiles that we probe in this work by neutrino oscillations in the presence of NSI. The neutrino oscillation probability differences governed by matter effects with NSI for different Earth density models are discussed in section~\ref{sec:NSI_effect}. In section~\ref{sec:events}, we present the method for simulating neutrino events at the ICAL detector and present the distributions of the event differences of reconstructed muon events between the standard three-layered density profile and alternative profiles. The statistical analysis method is demonstrated in section~\ref{sec:statistical analysis}. In section~\ref{sec:results}, we present the sensitivity to validate a high-density core inside Earth and measure the $R_\text{CMB}$ radius in the presence of NSI. Finally, we summarize our findings of this study and conclude in section~\ref{sec:conclusion}. The appendix~\ref{app:Puu-NSI} presents the effects of NSI parameters on the $\nu_\mu \rightarrow \nu_\mu$ disappearance channel. In appendix~\ref{app:Peu-NSI}, we discuss the effects of NSI parameters on the $\nu_e \rightarrow \nu_\mu$ appearance channel. The $P(\nu_{\mu} \rightarrow \nu_\mu)$ survival probability oscillograms, calculated using Earth density profiles with and without a core, as well as profiles corresponding to smaller and larger CMB radius, are presented in the appendix~\ref{app:Puu_oscillograms}. In appendix~\ref{app:cmb_results}, we show the sensitivity curves that are used to obtain constraints on the position of CMB radius in the presence of NSI. 

\section{Formalism of NSI}
\label{sec:formalism_nsi}

At low energies CC and NC neutrino NSI can be described by effective four-fermion dimension-six operators as follows~\cite{Wolfenstein:1977ue}:
\begin{equation}
	\mathcal{L}_{\text{CC-NSI}} = -2\sqrt{2}G_F \sum_{\alpha,\beta,C,f,f^\prime}\varepsilon_{\alpha\beta}^{ff^\prime C} \left(\bar{\nu}_\alpha\gamma^\rho P_L l_\beta\right)\left(\bar{f^\prime}\gamma_\rho P_C f\right)\,, 
	\label{eq:cc_nsi}
\end{equation}
\begin{equation}
	\mathcal{L}_{\text{NC-NSI}} = -2\sqrt{2}G_F \sum_{\alpha,\beta,C,f}\varepsilon_{\alpha\beta}^{fC} \left(\bar{\nu}_\alpha\gamma^\rho P_L \nu_\beta\right)\left(\bar{f}\gamma_\rho P_C f\right)\,,
	\label{eq:nc_nsi}
\end{equation}
where, $G_F$ is Fermi constant and $P_C\in\left\{P_R, P_L\right\}$ represents the right and left chiral projection operators $P_{R,L}=\left(1\pm \gamma_5\right)/2$. In eq.~\ref{eq:cc_nsi}, the dimensionless parameters $\varepsilon_{\alpha\beta}^{ff^\prime C}$ denote the strength of CC-NSI between the leptons of flavors $\alpha$ and $\beta$ ($\alpha$, $\beta$ = $e, \mu, \tau$), and matter fields $f\neq f^\prime \in \left\{u, d\right\}$. The dimensionless parameters $\varepsilon_{\alpha\beta}^{fC}$ in eq.~\ref{eq:nc_nsi} represent the strength of NC-NSI between the leptons of $\alpha$ and $\beta$ flavors ($\alpha$, $\beta$ = $e, \mu, \tau$), and the matter fields $f \in \left\{e, u, d\right\}$. The hermiticity of these interaction demands the following conditions:
\begin{equation}
	\varepsilon_{\alpha\beta}^{ff^\prime C} = \left(\varepsilon_{\beta\alpha}^{ff^\prime C}\right)^\ast ,  \quad	\varepsilon_{\alpha\beta}^{fC} = \left(\varepsilon_{\beta\alpha}^{fC}\right)^\ast \,. 
	\label{eq:hermicity_nsi}
\end{equation}

In this work, we study the impact of NC-NSI on atmospheric neutrinos, which pass long distances inside Earth and carry information about the internal structure of Earth through neutrino oscillations. The effective matter-induced potential generated by NC-NSI with all the matter fermions is 
\begin{equation}	\varepsilon_{\alpha\beta} \equiv \sum_{f = e, u, d} \left(\varepsilon_{\alpha\beta}^{fL} + \varepsilon_{\alpha\beta}^{fR} \right) \frac{V_f}{V_\text{CC}} \equiv \sum_{f = e, u, d}  \varepsilon_{\alpha\beta}^{f} \frac{N_f}{N_e} \,,
	\label{eq:nc_nsi2}
\end{equation}
where, $V_f = \sqrt{2}G_F N_f$, $V_\text{CC} = \sqrt{2}G_F N_e$, and $f = e,u,d$. The quantities $N_f$ and $N_e$ denote the number density of fermions ($e,u,d$) and electrons in the ambient medium, respectively. For antineutrino, $V_f \rightarrow - V_f$ and $V_\text{CC}\rightarrow - V_\text{CC}$. In this analysis, we consider Earth to be neutral and isoscalar, i.e. $N_n \approx N_p = N_e$, which leads to $N_u \approx N_d \approx 3N_e$. Thus, the effective NSI parameter,
\begin{equation}
\varepsilon_{\alpha\beta} \approx \varepsilon_{\alpha\beta}^e + 3\varepsilon_{\alpha\beta}^u + 3\varepsilon_{\alpha\beta}^d \, .
\label{eq:eps_convention}
\end{equation}
As the neutrino NSI with matter give rise to an additional matter potential apart from the standard MSW potential, the modified effective Hamiltonian in the presence of all possible NC-NSI of neutrino with matter fermions can be written as:
\begin{equation}
\begin{split}
H_{\text{eff}} &= \frac{1}{2E} \left[ U \begin{pmatrix}
0 & 0 & 0 \\
0 & \ms & 0 \\
0 & 0 & \ma
\end{pmatrix} U^\dagger+2EV_{CC}\begin{pmatrix}
1+\eee & \eem & \eet \\
\eems & \emm & \emt \\
\eets & \emts & \ett
\end{pmatrix} \right] \,,
\end{split}
\label{eq:hamiltonian}
\end{equation}
where, $\Delta m^2_{ij} \equiv m^2_i - m^2_j$ are the mass square differences, $U$ is the $3\times 3$ unitary Pontecorvo-Maki-Nakagawa-Sakata (PMNS) matrix which parametrizes neutrino mixing. The quantity $V_\text{CC}$ is the standard matter potential, which can be given as:
\begin{equation}
	V_\text{CC} = \sqrt{2} G_F N_e \approx \pm \, 7.6 \times Y_e \times 10^{-14} \left[\frac{\rho}{\text{g/cm}^3}\right]~\text{eV}\,,
\end{equation}
where, $Y_e = N_e/(N_p + N_n)$ is the relative electron number density inside the matter having density $\rho$. Further, $N_p$ and $N_n$ denote the number densities of protons and neutrons inside matter. Under the assumption that Earth is neutral and isoscalar, the electron-to-nucleon fraction is approximately $Y_e \approx 0.5$. From eq.~\ref{eq:hamiltonian}, the effective potential due to neutrino NSI would be $(V_\text{NSI})_{\alpha\beta} = \sqrt{2}G_F N_e \varepsilon_{\alpha\beta}$. For antineutrino, $U \rightarrow U^\ast$ and $\varepsilon_{\alpha\beta} \rightarrow \varepsilon^\ast_{\alpha\beta}$.

\begin{table}[htb!]
	\centering
	\begin{tabular}{|c|c|c|c|}
		\hline \hline
		\multirow{2}{*}{Experiment} & \multicolumn{3}{c|}{90\%  confidence level bounds} \\ 
		\cline{2-4}
		& $\varepsilon_{e\mu}$ & $\varepsilon_{e\tau}$ & $\varepsilon_{\mu\tau}$ \\ \hline
		IceCube~\cite{IceCube:2022ubv} & - & - & $-0.012 <\varepsilon_{\mu\tau}< 0.009$ \\ \hline
		DeepCore~\cite{IceCubeCollaboration:2021euf} & $|\varepsilon_{e\mu}| \leq 0.143$ & $|\varepsilon_{e\tau}| \leq 0.17$ & $|\varepsilon_{\mu\tau}| \leq 0.0228$ \\ \hline
		ANTARES~\cite{ANTARES:2021crm} & - & - & $-1.4 \times 10^{-2} < \varepsilon_{\mu\tau} < 0.87 \times 10^{-2}$ \\ \hline
		KM3NET/ORCA~\cite{KM3NeT:2024pte} & $|\varepsilon_{e\mu}| \leq 0.168$ & $|\varepsilon_{e\tau}| \leq 0.22$ & $|\varepsilon_{\mu\tau}| \leq 0.0162$ \\ \hline
		NOvA~\cite{NOvA:2024lti} & $|\varepsilon_{e\mu}| \leq 0.3$ & $|\varepsilon_{e\tau}| \leq 0.4$ & - \\ \hline
	\end{tabular}
	\mycaption{The  90\% confidence level bounds on the NC-NSI parameters $\varepsilon_{e\mu}$, $\varepsilon_{e\tau}$, and $\varepsilon_{\mu\tau}$ from various experiments. The bounds have been converted to the convention used in the present work (see eq.~\ref{eq:eps_convention}).}
	\label{tab:nsi_bounds}
\end{table}

Table~\ref{tab:nsi_bounds} summarises the current bounds on the flavor-violating NC-NSI parameters at 90\% confidence level from various experiments. Among all the NSI parameters, the $\varepsilon_{\mu\tau}$ is having most stringent constraint from the IceCube experiment~\cite{IceCube:2022ubv}.

\section{Earth density profiles}
\label{sec:earth-profile}

\begin{figure}[htb!]
    \centering
    \includegraphics[width=0.65\linewidth]{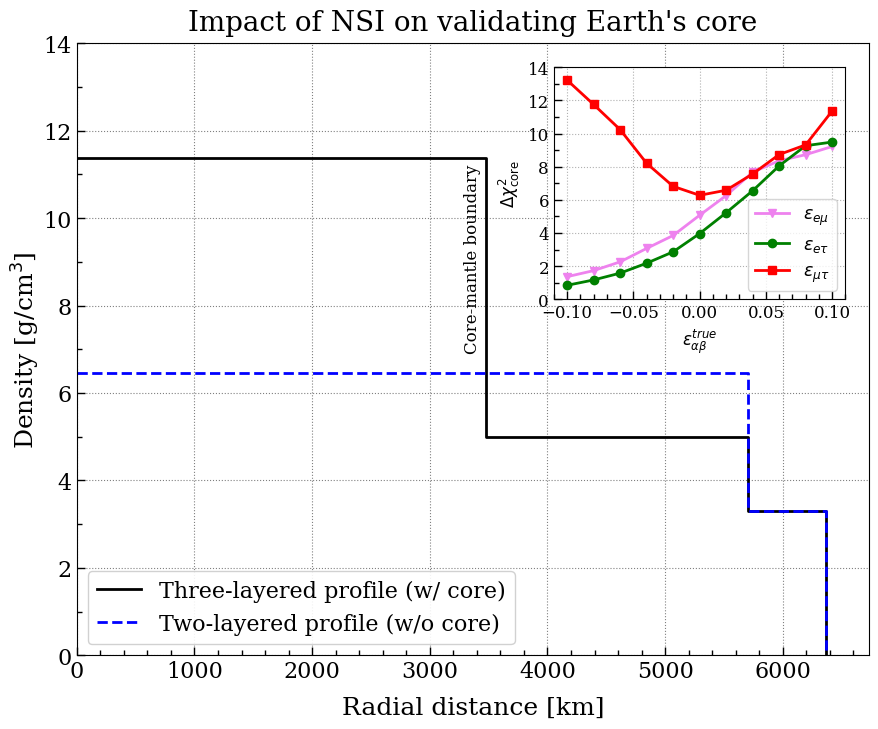}
    \mycaption{Layer densities as functions of radial distances for the three-layered and two-layered density profiles of Earth. The solid-black curve represents the density profile with a core, whereas the dashed-blue curve corresponds to a density profile without a core. The inset plot illustrates the expected sensitivity to validate the core inside Earth in the presence of NC-NSI parameters. See section~\ref{sec:results_vc} for details. {\it\textbf{Our sensitivities indicate that the significance for validating the core is substantially affected by the presence of NSI.}}}
    \label{fig:dens_profile_core}
\end{figure}

\begin{figure}[htb!]
    \centering
    \includegraphics[width=1.0\linewidth]{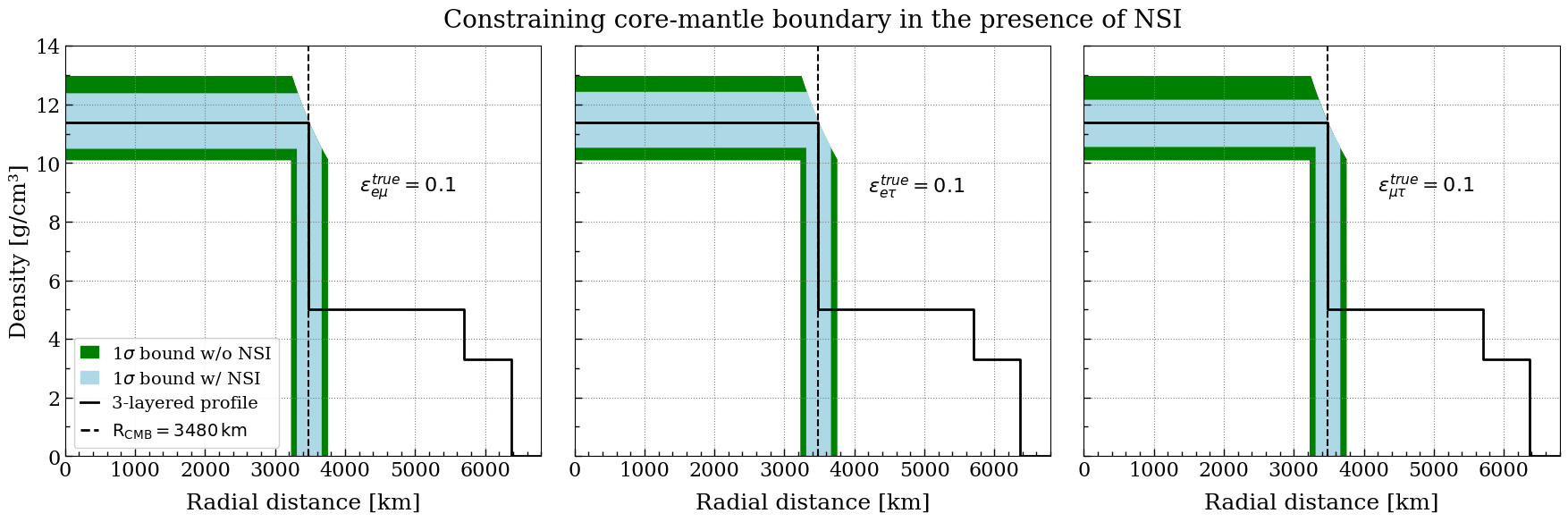}
    \mycaption{Layer densities as functions of radial distances for the three-layered density profiles of Earth with modified-CMB locations. The solid-black curves represent the standard three-layered density profile. The dashed-black lines indicate the standard CMB location. The dark-green (light-blue) bands in all three panels show the $1\sigma$ bounds on the CMB location with SI (NSI). The dark-colored bands in the left, middle, and right panels correspond to the NC-NSI parameters $\varepsilon_{e\mu}$, $\varepsilon_{e\tau}$, and $\varepsilon_{\mu\tau}$ with the true values of $0.1$, respectively. See section~\ref{sec:results_cmb} for details. {\it\textbf{These plots demonstrate that the limits on the location of the core-mantle boundary are significantly modified by the presence of NSI.}}}
    \label{fig:dens_profile_cmb}
\end{figure}

In this study, we consider a simple three-layered density profile guided by the PREM~\cite{Dziewonski:1981xy} as the standard density model of Earth. In this profile, the boundaries of the three layers are assumed to be at the same locations as in the PREM, whereas the densities of these three layers are calculated by merging finer PREM layers within them and then averaging the densities of finer layers. The solid black curves in figure~\ref{fig:dens_profile_core} and figure~\ref{fig:dens_profile_cmb} present this standard three-layered density profile, where a significant density discontinuity occurs at the CMB with radius $R_\text{CMB} = 3480$ km or depth of 2891 km. We refer to this profile as a density profile with a core, which is abbreviated as ``w/ core". The dashed blue curve shown in figure~\ref{fig:dens_profile_core} represents a two-layered density profile of Earth without a core, which is abbreviated as ``w/o core''. In this two-layered profile, the core and inner mantle are merged to form a modified inner mantle such that the total mass of Earth remains consistent with that of the standard three-layered profile. In these two- and three-layered density models, the moment of inertia of Earth is not imposed as a constraint. These simplified models may have slight deviations compared to realistic Earth models, however they are efficient in demonstrating the analysis methodology where neutrinos can be used to probe the broad features of Earth. In this work, we aim to quantify the expected sensitivity of the ICAL detector to differentiate between these two density models using atmospheric neutrino oscillations in the presence of NSI. The inset plot in figure~\ref{fig:dens_profile_core} illustrates the expected sensitivity to validate the presence of core by ruling out the density profile without a core with respect to the profile with a core, expressed in terms of $\Delta\chi^2_\text{core}$ (as defined later in eq.~\ref{eq:chi2_vc}) as a function of the true values of the NSI parameters. The magenta, green, and red curves represent the sensitivity in the presence of NC-NSI parameters $\varepsilon_{e\mu}$, $\varepsilon_{e\tau}$, and $\varepsilon_{\mu\tau}$, respectively. A detailed description of this inset plot is presented in section~\ref{sec:results_vc}.

Figure~\ref{fig:dens_profile_cmb} illustrates the three-layered density profiles of Earth with different modified-CMB locations. We modify the location of CMB with respect to its standard value of $R_\text{CMB}$ (standard) = 3480 km. The CMB radius smaller (larger) than the standard CMB is referred to as a smaller (larger) core or SC (LC). The densities of inner and outer mantles remain fixed during the CMB modification. The density of the core is modified to conserve the mass of Earth. We aim to constrain the CMB location in the presence of NC-NSI parameters. To demonstrate it, we present the $1\sigma$ bound on the $R_\text{CMB}$ radius with a representative choice of NC-NSI parameters $\varepsilon^{\text{true}}_{\alpha\beta} = 0.1$. The light-blue bands in the left, middle, and right panels of figure~\ref{fig:dens_profile_cmb} show the $1\sigma$ bounds on the CMB location in the presence of $\varepsilon^{\text{true}}_{e\mu} = 0.1$, $\varepsilon^{\text{true}}_{e\tau} = 0.1$, and $\varepsilon^{\text{true}}_{\mu\tau} = 0.1$, respectively. For a comparison, the dark-green bands in each panel present $1\sigma$ bound without NSI. See section~\ref{sec:results_cmb} for more details.

\section{Impact of NSI on neutrino oscillations}
\label{sec:NSI_effect}

In this section, we discuss how the oscillation patterns of the upward-going atmospheric neutrinos get affected in the presence of NSI. Atmospheric neutrinos are produced during the interactions of cosmic ray particles with the air nuclei in the Earth's atmosphere. Their flux predominantly comprises electron and muon flavors. These neutrinos traverse a wide range of baselines, from approximately 15 km to 12757 km, and span an energy spectrum from several MeV to tens of TeV. The upward-going multi-GeV atmospheric neutrinos experience the Earth matter effects, which modify the neutrino oscillation probabilities. In the presence of NSI, neutrinos encounter an additional matter potential beyond the standard matter potential, further altering the neutrino oscillation probabilities.

\begin{table}
\centering
\begin{tabular}{|c|c|c|c|c|c|c|}
\hline
$\sin^2 2\theta_{12}$ & $\sin^2\theta_{23}$ & $\sin^2 2\theta_{13}$ & $\Delta m^2_
\text{eff}$ (eV$^2$) & $\Delta m^2_{21}$ (eV$^2$) & $\delta_{\rm CP}$ & Mass Ordering\\
\hline
0.855 & 0.5 & 0.0875 & $2.49\times 10^{-3}$ & $7.4\times10^{-5}$ & 0 & Normal (NO)\\
\hline 
\end{tabular}
\mycaption{The benchmark values of oscillation parameters considered in this analysis. These values are in good agreement with the present neutrino global fits~\cite{Capozzi:2025wyn,Esteban:2024eli,NuFIT,Capozzi:2021fjo,deSalas:2020pgw}.}
\label{tab:osc-param-value}
\end{table}

In the present work, we numerically compute neutrino oscillation probabilities within the three-flavor paradigm for various $L \text{(km)}/E \text{(GeV)}$ combinations available for atmospheric neutrinos. We utilize different Earth density profiles and compare the resultant neutrino oscillation probabilities for each profile. We employ benchmark values of neutrino oscillation parameters as given in table~\ref{tab:osc-param-value}. The value of $\Delta m^2_{31}$ is derived from the effective atmospheric mass-squared difference\footnote{The effective atmospheric mass-squared difference is defined in terms of $\Delta m^2_{31}$ and $\Delta m^2_{21}$ as follows~\cite{deGouvea:2005hk,Nunokawa:2005nx}:
	\begin{equation}
		\Delta m^2_\text{eff} = \Delta m^2_{31} - \Delta m^2_{21} (\cos^2\theta_{12} - \cos \delta_\text{CP} \sin\theta_{13}\sin2\theta_{12}\tan\theta_{23}) \,.
		\label{eq:m_eff}
\end{equation}} $\Delta m^2_{\text{eff}}$. The positive value of $\Delta m^2_{\text{eff}}$ is taken when we consider the neutrino mass ordering to be normal (NO), whereas the negative value with the same magnitude is taken when we consider the neutrino mass ordering to be inverted (IO).

As upward-going atmospheric neutrinos traverse Earth, their effective masses and flavor mixing are modified due to the standard matter potential~\cite{Wolfenstein:1977ue,Mikheev:1986gs,Mikheev:1986wj} as well as the NSI, consequently altering neutrino oscillation probabilities. The matter effect caused by the standard matter potential is significant for neutrinos (antineutrinos) if the neutrino mass ordering is NO (IO). These resonant enhancements depend upon the energy of neutrinos and the density distribution of electrons they encounter while passing through Earth. In particular, the mantle-passing neutrinos with energies around $6-10$ GeV experience the MSW resonance~\cite{Wolfenstein:1977ue,Mikheev:1986gs,Mikheev:1986wj}. Furthermore, the core-passing neutrinos encounter a sharp density transition at the CMB, resulting in the additional matter effect due to the parametric resonance (PR)~\cite{Ermilova:1986,Akhmedov:1988kd,Krastev:1989,Akhmedov:1998ui,Akhmedov:1998xq} or neutrino oscillation length resonance (NOLR)~\cite{Petcov:1998su,Chizhov:1998ug,Petcov:1998sg,Chizhov:1999az,Chizhov:1999he} around $3-6$ GeV. The PR/NOLR also depends upon the magnitude of the density transition at the CMB and its location. Therefore, any modification in the density jump at the CMB or its location alters the PR/NOLR resonance and, hence, the oscillation probabilities of neutrinos passing through the core.

\begin{figure}[htb!]
\centering
\includegraphics[width=1.0\linewidth]{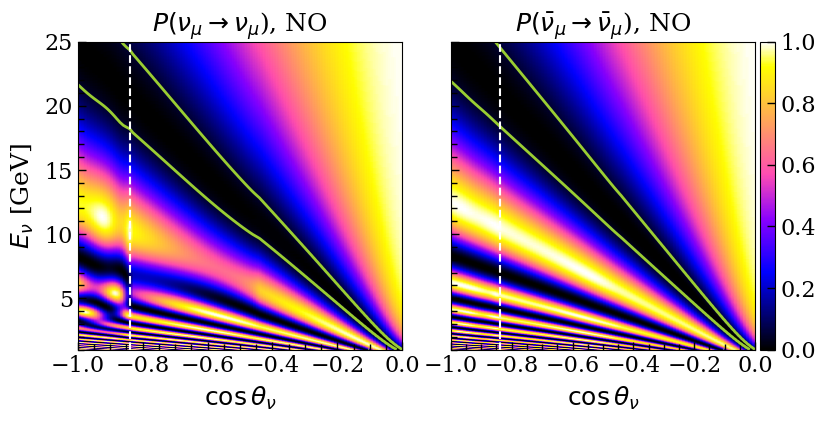}
\mycaption{The three-flavor $\nu_\mu \rightarrow \nu_\mu$ survival probability oscillogram for the standard three-layered density profile of Earth with standard CMB radius, $R_\text{CMB} = 3480$ km (as shown by the vertical dashed-white lines). The left (right) panel corresponds to neutrino (antineutrino). The green bands represent the first oscillation valley. We consider the benchmark values of neutrino oscillation parameters as given in table~\ref{tab:osc-param-value}, assuming the neutrino mass ordering to be NO.}
\label{fig:sm-oscillogram}
\end{figure}

The neutrino events at ICAL would be contributed by the $\nu_\mu \rightarrow \nu_\mu$ disappearance channel and the $\nu_e \rightarrow \nu_\mu$  appearance channel. However, more than 98\% of the muon-type neutrino events are contributed by the $\nu_\mu \rightarrow \nu_\mu$ disappearance channel. Therefore, in this section, we primarily present the effect of the Earth density profile with (without) a core and the modification of CMB location in the presence of  NC-NSI parameters only for the $\nu_\mu \rightarrow \nu_\mu$ disappearance probability channel. Figure~\ref{fig:sm-oscillogram} presents the three-flavor P($\nu_\mu \rightarrow \nu_\mu$) survival probability oscillograms in the plane of $(E_\nu, \cos\theta_\nu)$ for neutrinos (left panel) and antineutrinos (right panel) assuming NO. We consider the standard three-layered density profile of Earth while making these oscillograms. In both panels, the neutrino energy range is from $1-25$ GeV and the cosine of zenith angle $\cos\theta_\nu$ is from -1 to 0. The dark-diagonal band highlighted by the green curves, extending from ($E_\nu = 1$ GeV, $\cos\theta_\nu = 0$) to ($E_\nu = 25$ GeV, $\cos\theta_\nu = -\,1$), represents the first oscillation minimum, which is also known as the ``oscillation valley''~\cite{Kumar:2020wgz,Kumar:2021lrn}. In the left panel, we can observe the distortion in the oscillation probability due to the standard matter effects around $-\,0.8 < \cos\theta_\nu < -\,0.5$ and 6 GeV $< E_\nu <$ 10 GeV, which corresponds to the MSW resonance. Another distortion can be observed around $\cos\theta_\nu< -\,0.8$  and 3 GeV $< E_\nu <$ 6 GeV, which is due to the PR/NOLR resonance. In the right panel, these matter-effect regions are not observed for the antineutrinos as the considered mass ordering is NO, which is similar to the vacuum oscillations. However, this trend reverses for inverted mass ordering, where antineutrinos experience a significant amount of matter effects while neutrinos do not. Till now, we have discussed the impact of the standard matter potential on P$(\nu_\mu \rightarrow \nu_\mu)$ survival probabilities, the impact of flavor-violating NC-NSI parameters on neutrino oscillation patterns is discussed in appendix~\ref{app:Puu_oscillograms}.

\subsection{Impact of NSI on oscillograms for Earth density profile w/ and w/o core}
\label{sec:oscillograms_vc}

\begin{figure}[htb!]
	\centering
	\includegraphics[width=1.0\linewidth]{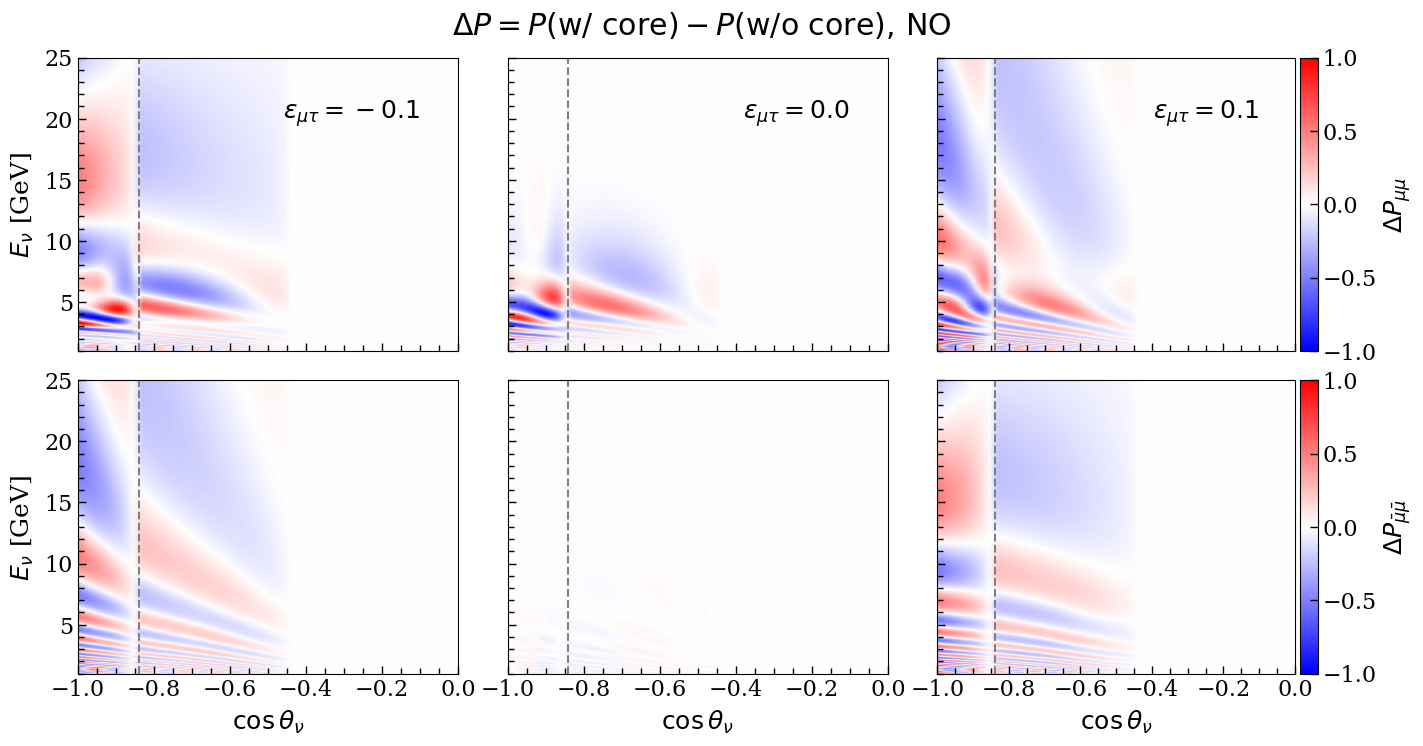}
	\mycaption{The top (bottom) panels show differences between the three-flavor neutrino oscillation probabilities for ${\nu_\mu} \rightarrow {\nu_\mu}$ ($\bar{\nu}_{\mu} \rightarrow \bar{\nu}_{\mu}$) channel considering Earth density models with and without a core. The left, middle, and right columns correspond to the probability differences with NC-NSI parameter $\varepsilon_{\mu\tau}=-\,0.1, \,0.0,$ and 0.1, respectively. The top and bottom middle columns with $\varepsilon_{\mu\tau} = 0.0$ represent the SI scenario. The dashed-gray vertical lines in each panel represent the standard CMB position with radius of $R_\text{CMB} = 3480$ km. We consider the benchmark values of neutrinos oscillation parameters as given in table~\ref{tab:osc-param-value} assuming NO.} 
	\label{fig:vc-mutau}
\end{figure}

In figure~\ref{fig:app-vc-mumu} of appendix~\ref{app:Puu_oscillograms}, we present the effect of the three-layered (w/ core) and the two-layered (w/o core) density models on neutrino oscillations with NSI. In this section, we now show the neutrino oscillation probability differences between these two Earth density profiles in the presence of flavor-violating NC-NSI parameters. The top (bottom) panels of figure~\ref{fig:vc-mutau} show the difference between $\nu_\mu$ ($\bar{\nu}_\mu$) survival probabilities for density profiles with core and without core in the presence of NC-NSI parameter $\varepsilon_{\mu\tau}$ assuming the mass ordering to be NO. These probability differences are defined as:
\begin{equation}
\Delta P_{\mu\mu} = P(\nu_\mu \rightarrow \nu_\mu) \, \text{[w/ core]} -  P(\nu_\mu \rightarrow \nu_\mu) \, \text{[w/o core]} \, , 
\label{eq:puu_prob_diff}
\end{equation}
\begin{equation}
\Delta P_{\bar{\mu}\bar{\mu}} = P(\bar{\nu}_\mu \rightarrow \bar{\nu}_\mu) \, \text{[w/ core]} -  P(\bar{\nu}_\mu \rightarrow \bar{\nu}_\mu) \, \text{[w/o core]} \, .
\label{eq:pauu_prob_diff}
\end{equation}
The left, middle, and right panels correspond to the oscillation probability differences for $\varepsilon_{\mu\tau}=-\,0.1$, $\varepsilon_{\mu\tau}=0.0$ (or SI), and $\varepsilon_{\mu\tau}=0.1$, respectively. From the top middle panel of figure~\ref{fig:vc-mutau}, we can observe that the probability differences are non-zero only at the higher baselines ($\cos\theta_\nu < -\,0.6$) and lower energies ($E_\nu < 10$ GeV) where the standard matter effects are prominent. In particular, the differences are significant for the core-passing trajectories due to the absence of the core in the two-layered density profile. In the bottom middle panel, the probability differences vanish for the $\bar{\nu}_\mu$ survival probability because the standard matter effects are not significant for antineutrino oscillations if the mass ordering is NO.

From the top left (right) panel of figure~\ref{fig:vc-mutau}, we can observe that the probability differences are significant even for higher energies in the presence of NC-NSI parameter $\varepsilon_{\mu\tau}=-\,0.1$ ($\varepsilon_{\mu\tau}=0.1$) compared to the top middle plot. The probability differences are now also present for $\bar{\nu}_\mu$ survival probabilities with NSI, as seen from the bottom left and right panels of figure~\ref{fig:vc-mutau}. The probability differences are opposite to each other for positive and negative value of $\varepsilon_{\mu\tau}$. Further, the $\nu_\mu$ survival probability differences for positive (negative) value of $\varepsilon_{\mu\tau}$ are identical to the $\bar{\nu}_\mu$ survival probability differences for negative (positive) value of $\varepsilon_{\mu\tau}$. All these effects of $\varepsilon_{\mu\tau}$ on $\nu_\mu$ and $\bar{\nu}_\mu$ survival probability differences can be explained by the eq.~\ref{eq:Pmuu-mat-NSI} in appendix~\ref{app:Puu-NSI}. The left and right panels of figure~\ref{fig:vc-mutau} indicate that we can differentiate between the density profile with core and without core more significantly in the presence of NC-NSI parameter $\varepsilon_{\mu\tau}$ as compared to the SI scenario, even in the case of antineutrinos assuming NO.

\begin{figure}[htb!]
	\centering
	\includegraphics[width=1.0\linewidth]{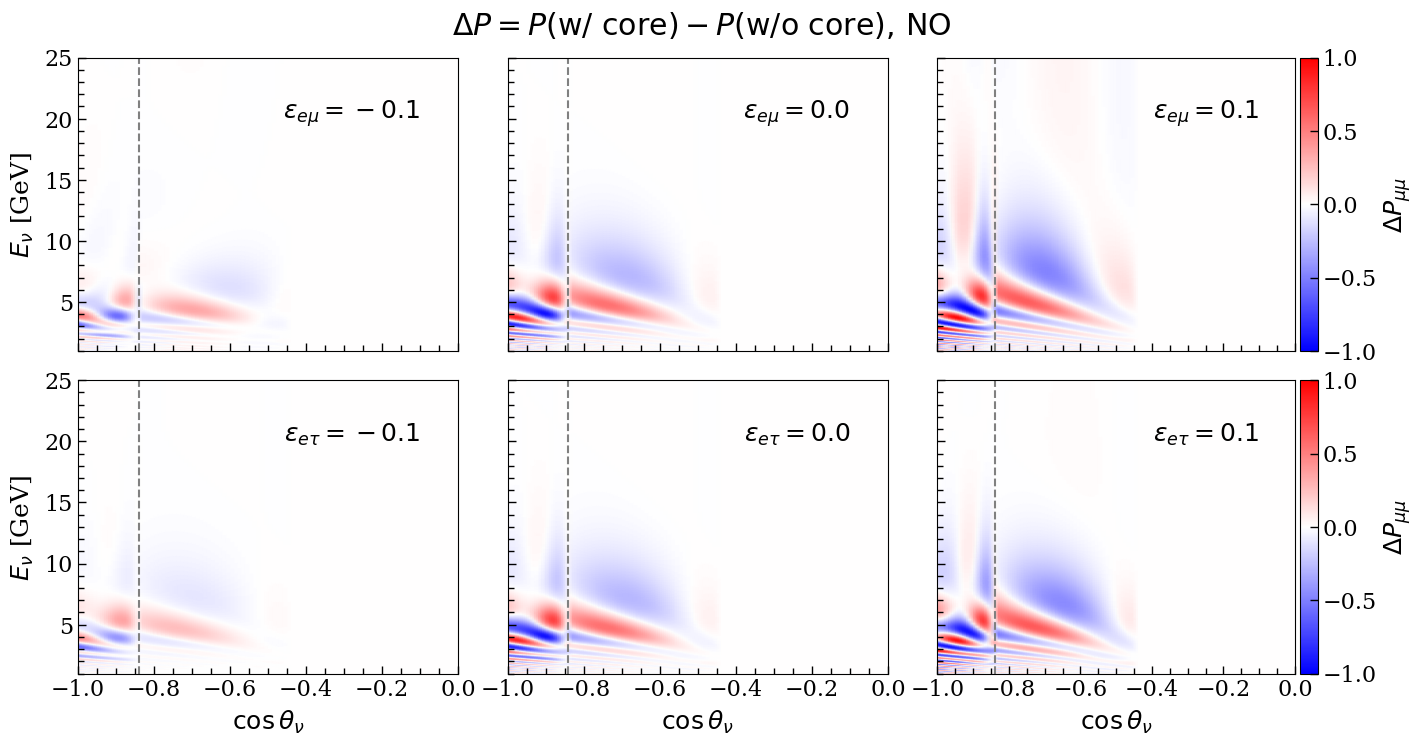}
	\mycaption{Difference in the three-flavour neutrino oscillation probabilities for ${\nu_\mu} \rightarrow {\nu_\mu}$ channel between the Earth density profiles with and without core in the presence of NC-NSI parameters $\varepsilon_{e\mu}$ (top panel) and $\varepsilon_{e\tau}$ (bottom panel). The left, middle, and right columns correspond to the probability differences for value of $\varepsilon_{e\mu}$ (or  $\varepsilon_{e\tau}) = $ $-\,0.1$, $0.0$, and $0.1$, respectively. The middle column presents the SI scenario. The dashed-grey vertical lines in each panel represent the standard CMB position with radius of $R_\text{CMB} = 3480$ km. We consider the benchmark oscillation parameters as given in table~\ref{tab:osc-param-value}, assuming NO.}
	\label{fig:vc-emu-etau}
\end{figure}

In the top (bottom) row of figure~\ref{fig:vc-emu-etau}, we present the impact of NC-NSI parameter $\varepsilon_{e\mu}$ ($\varepsilon_{e\tau}$) on ${\nu_\mu} \rightarrow {\nu_\mu}$ survival probability differences in the plane of neutrino energy $(E_\nu)$ and direction $(\cos\theta_\nu)$ between the density profiles of Earth with core and without core assuming NO. The probability differences are the same as defined in eq.~\ref{eq:puu_prob_diff}. We consider three different values of NC-NSI parameters $\varepsilon_{e\mu}$ and $\varepsilon_{e\tau}$ which are $-\,0.1$, $0$, and $0.1$ as shown in the left, middle, and right columns, respectively. The probability differences for the top and bottom middle panels are similar to the top middle panel of figure~\ref{fig:vc-mutau}. Moreover, the probability differences are less (more) pronounced for negative (positive) values of these NC-NSI parameters compared to the case of SI. These effects of $\varepsilon_{e\mu}$ and $\varepsilon_{e\tau}$ on $P(\nu_\mu \rightarrow \nu_\mu)$ survival probability can be explained as follows.

The effect of $\varepsilon_{e\mu}$ and $\varepsilon_{e\tau}$ emerge at subleading order in the $P({\nu_\mu} \rightarrow {\nu_\mu})$ disappearance probability expression, which is non-trivial to describe analytically. However, their effects can be analyzed by studying the analytical expression for $P({\nu_e} \rightarrow {\nu_\mu})$ appearance probability as provided in eq.~\ref{eq:Pmue-mat-NSI} in appendix~\ref{app:Peu-NSI}, where the impacts of $\varepsilon_{e\mu}$ and $\varepsilon_{e\tau}$ appear in the leading order terms. The effect of NSI parameter $\varepsilon_{e\mu}$ is primarily contributed by the fifth term in eq.~\ref{eq:Pmue-mat-NSI}, which is expressed as:
\begin{equation}
	+ 8 \epsilon_{e\mu} \tilde{s}_{13} s_{23}^{3} \frac{a_{\rm CC}}{\ldm - a_{\rm CC}}\sin^{2} \frac{(\ldm - a_{\rm CC})L}{4E} \, ,
\end{equation}
where, the factor ($\ldm - a_{\rm CC}$) in the denominator induces the standard matter-driven resonance effects for neutrinos (antineutrinos) if the neutrino mass ordering is normal (inverted). Consequently,  a positive (negative) value of $\varepsilon_{e\mu}$ increases (decreases) $P(\nu_e \rightarrow \nu_\mu)$ for neutrinos, which can be translated as a
decrease (increase) in $P(\nu_\mu \rightarrow \nu_\mu)$ survival probability because we have,
\begin{equation}
	P(\nu_\mu \rightarrow \nu_\mu) = 1 - P(\nu_\mu \rightarrow \nu_e) - P(\nu_\mu \rightarrow \nu_\tau) \,,
\end{equation}
where, $P(\nu_\mu \rightarrow \nu_e) = P(\nu_e \rightarrow \nu_\mu)$ for $\delta_{\rm CP} = 0$. It is important to note that the $\nu_\mu \rightarrow \nu_\tau$ oscillation channel is also influenced by the matter effects for certain energy and baseline ranges~\cite{Gandhi:2004md}. However, the impact of the NSI parameter $\varepsilon_{e\mu}$ in this channel is limited, as it appears only at the subleading order. Further, an equivalent effect is also observed for NC-NSI parameter $\varepsilon_{e\tau}$, which can be understood through the seventh term in eq.~\ref{eq:Pmue-mat-NSI}.

The effect of NC-NSI parameters $\varepsilon_{e\mu}$ and $\varepsilon_{e\tau}$ are significant in the case of neutrinos as compared to antineutrinos. This happens because $a_{\rm CC}$ becomes negative for antineutrinos, preventing the matter-driven resonance condition from being satisfied for the NO. As a result, the aforementioned term contributes less significantly to the oscillation probabilities for antineutrinos. That is why, we have not presented the effect of $\varepsilon_{e\mu}$ and $\varepsilon_{e\tau}$ on $\bar{\nu}_\mu$ survival probability like the effect of $\varepsilon_{\mu\tau}$.

The above-discussed effects for the NC-NSI parameters are also evident in our sensitivity results where we present the sensitivity for validating the density profile with core by ruling out the density profile without core using a magnetized ICAL detector.

\subsection{Impact of NSI on oscillograms for modified core-mantle boundary location}
\label{sec:oscillograms_cmb}

\begin{figure}[htb!]
	\includegraphics[width=1.0\linewidth]{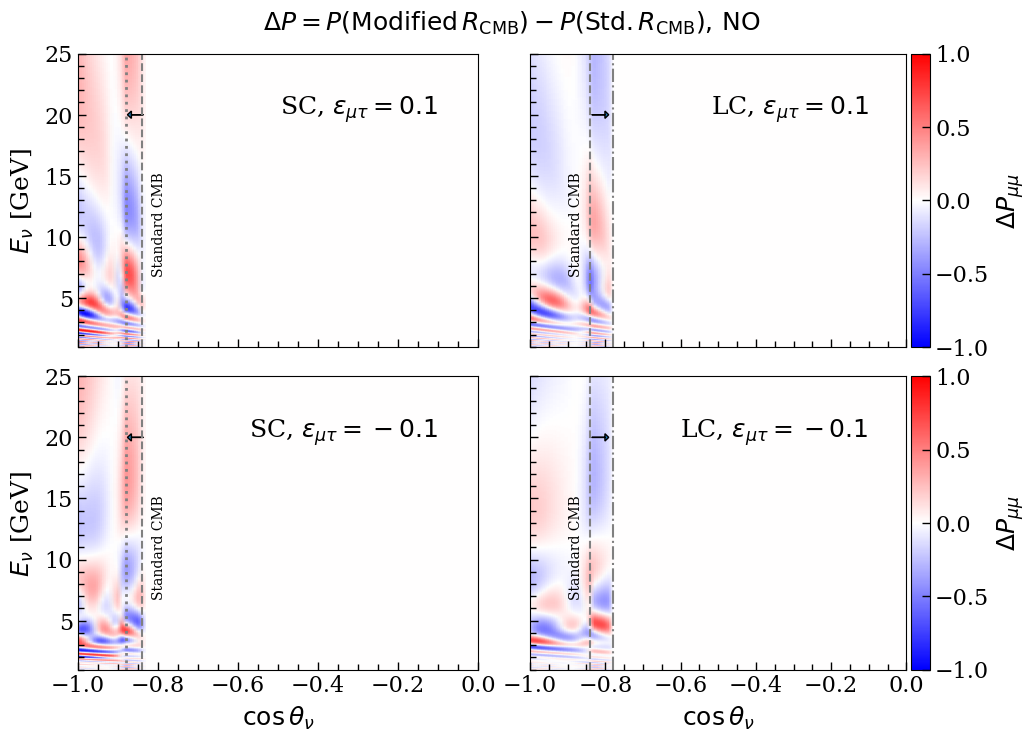}
	\mycaption{Differences in the three-flavour neutrino oscillation probabilities for the $\nu_{\mu} \rightarrow \nu_{\mu}$ channel with the modified $R_\text{CMB}$ and the standard $R_\text{CMB}$ in the presence of the NC-NSI parameter $\varepsilon_{\mu\tau}$. The top (bottom) row corresponds to $\varepsilon_{\mu\tau} = 0.1$ ($-\,0.1$). The left and the right columns correspond to the smaller ($-\,500$ km) and the larger ($+\,500$ km) values of $R_\text{CMB}$, respectively, where the standard $R_\text{CMB} = 3480$ km. The dashed, dotted, and dot-dashed gray vertical lines in each panel represent the standard CMB, SC, and LC. The black arrows demonstrate the direction of the CMB modification from its standard location. We consider the benchmark oscillation parameters as given in table~\ref{tab:osc-param-value} assuming NO.}
	\label{fig:cmb-mutau}
\end{figure}

The effects of the Earth density profile with the modified-CMB location on neutrino oscillation probabilities with NSI are presented in figure~\ref{fig:app-cmb-numu} of appendix~\ref{app:Puu_oscillograms}. In this section, we now discuss the neutrino oscillation probability differences between the density profiles with the modified-CMB location and the standard CMB location in the presence of flavor-violating NC-NSI parameters. Figure~\ref{fig:cmb-mutau} presents the difference between $\nu_\mu$ survival probability for the SC or LC scenario and the standard $R_\text{CMB}$  in the presence of NC-NSI parameter $\varepsilon_{\mu\tau}$ with the value of $0.1$ (top panel) and $-\,0.1$ (bottom panel). The left (right) column corresponds to probability differences between SC (LC) and standard $R_\text{CMB}$ scenario. We consider a modification in the CMB location by $-\,500$ km ($+\,500$ km) for SC (LC) scenarios from its standard value.

For the case of the standard interactions, it was observed that the probability differences are non-zero only in the core regions, while they are not present in the mantle region because the density of the mantle remains the same for the modified and standard CMB scenarios as described by the middle row of figure~6 in Ref.~\cite{Upadhyay:2022jfd}. 

From figure~\ref{fig:cmb-mutau}, we can observe that in the presence of flavor-violating NC-NSI parameter $\varepsilon_{\mu\tau}$, the probability differences are extended to the higher energies as compared to the SI case. The effects of $\varepsilon_{\mu\tau}$ on $P(\nu_\mu \rightarrow \nu_\mu)$ survival probability can be explain by the eq.~\ref{eq:Pmuu-mat-NSI} in appendix~\ref{app:Puu-NSI}.

\section{Atmospheric neutrino event generation at ICAL}
\label{sec:events}
The iron calorimeter is a 50 kton magnetized detector at the proposed India-based Neutrino Observatory~\cite{ICAL:2015stm}. The ICAL detector is designed to detect atmospheric neutrinos and antineutrinos separately in the multi-GeV energy range, covering a wide range of baselines. It would have three modules of size $\text{16 m}\times \text{16 m}\times \text{14.5 m}$, each consisting of about 151 alternative layers of iron plates having a thickness of 5.6 cm. There would be a vertical gap of about 4 cm between two iron plates to accommodate the glass Resistive Plate Chambers (RPCs), which are $\text{2 m}\times \text{2 m}$ in size. The iron plates provide the target mass for neutrino interactions, enabling the production of secondary charged particles like muons and hadrons during CC interactions of neutrinos with iron nuclei. These secondary charged particles are detected by the RPCs. Thus, the iron plates act as passive detector elements, while the RPCs function as active detector elements. As charged particles propagate through the detector, they deposit energy in the form of hits in the RPCs. The pickup strips determine the X and Y coordinates of these hits, while the RPC layer number provides the Z coordinate.

In the multi-GeV energy range, neutrino interactions occur through the resonance and deep inelastic scatterings processes that produce leptons and hadrons.
A multi-GeV muon produced in the charged-current interaction of muon neutrino passes through multiple layers of the detector, leaving hits that form a track due to its minimum ionization. The ICAL detector, with a magnetic field of about 1.5~T~\cite{Behera:2014zca}, would distinguish neutrinos ($\nu_\mu$) from antineutrino ($\bar{\nu}_\mu$) by observing the curvature of the muon tracks. The magnetic field bends the paths of $\mu^-$ and $\mu^+$ in the opposite directions, allowing for clear differentiation between neutrinos ($\nu_\mu$) and antineutrinos ($\bar{\nu}_\mu$). This charge identification capability (CID) of the ICAL detector would play a crucial role in analyses driven by matter effects, such as determining the neutrino mass ordering, measuring the octant of $\theta_{23}$, and conducting neutrino oscillation tomography. Further, the nanosecond-level time resolution of RPCs~\cite{Dash:2014ifa,Bhatt:2016rek,Gaur:2017uaf} enables ICAL to distinguish between upward-going and downward-going muon events separately. As far as the hadrons produced in the neutrino interactions are concerned, they deposit their energies in the form of multiple hits in a given RPC layer and form a shower-like event. These hadrons carry a significant fraction of incoming neutrino energy ($E_\nu$), which is defined as ${E'}_\text{had} = E_\nu - E_\mu$ where $E_\mu$ is the amount of energy carried away by a muon. 

In the present study, we simulate the unoscillated neutrino events with the NUANCE~\cite{Casper:2002sd} Monte Carlo (MC) neutrino event generator, utilizing the geometry of ICAL as target and the Honda flux~\cite{Athar:2012it,Honda:2015fha} for atmospheric neutrinos at the proposed INO site at Theni district of Tamil Nadu, India. We generate the unoscillated MC neutrino events over a substantial exposure time of 1000 years to minimize the statistical fluctuations. The solar modulation effect on the atmospheric neutrino flux has been incorporated by considering flux during periods of high solar activity for half of the exposure time and low solar activity for the other half. At the INO site, the downward-going cosmic muon background would be suppressed by a factor of approximately $10^6$~\cite{Dash:2015blu} due to a mountain overburden of at least 1 km (3800 m water equivalent) from all directions. Additionally, the analysis at ICAL considers muon events with vertices completely inside the detector and far from the edges to exclude muon events entering from outside the detector. Consequently, we expect a negligible downward-going cosmic muon background at ICAL. We do not consider muon events which are produced by the interaction of tau neutrinos, as they constitute only 2\% of the total upward-going muons from $\nu_\mu$ interactions, and most of these events have energies below the detection threshold ($\sim 1$ GeV) of ICAL. We implement the three-flavor neutrino oscillations in the presence of Earth's matter effects using a reweighting algorithm~\cite{Devi:2014yaa,Ghosh:2012px,Thakore:2013xqa}.

The detector responses for muons~\cite{Chatterjee:2014vta} and hadrons~\cite{Devi:2013wxa} have been implemented through the migration matrices developed by the ICAL collaboration, based on an extensive detector simulation study performed with the GEANT4~\cite{Geant4:2003} package. Figures 6, 11, 13, and 14 of ref.~\cite{Chatterjee:2014vta} present the angular resolution, energy resolution, reconstruction efficiency, and CID efficiency of the ICAL detector for reconstructed muons, respectively. In the energy range of $1-25$ GeV, ICAL would achieve an excellent angular resolution of approximately $1^\circ$~\cite{Chatterjee:2014vta} for measuring the directions of muon ($\cos\theta_\mu$) and a resolution of about 10\% to 15\%~\cite{Chatterjee:2014vta} for determining the energy of muon ($E_\mu$). For hadrons with energies (${E'}_\text{had}$) above 5 GeV, ICAL provides an energy resolution of about 40\%~\cite{Devi:2013wxa}. After applying the detector response, we derive the reconstructed observables for the muon energy ($E_\mu^\text{rec}$), the muon direction ($\cos\theta_\mu^\text{rec}$), and the hadron energy (${E^\prime}_{\text{had}}^\text{rec}$).

\subsection{Total event rates}

\begin{table}[htb!]
	\centering
	\begin{tabular}{| c | c | c |  c | c | c | c | c | c |} 
		\hline \hline
		& \multicolumn{4}{c|}{Three-layered profile (w/ core)} & 
		\multicolumn{4}{c|}{Two-layered profile (w/o core)}  \\
		\cline{2-9}
		& \multicolumn{2}{c|}{$\mu^-$ events} & 
		\multicolumn{2}{c|}{$\mu^+$ events} &  \multicolumn{2}{c|}{$\mu^-$ events} & 
		\multicolumn{2}{c|}{$\mu^+$ events}  \\
		\cline{2-9}
		& -\,0.1 & +\,0.1 & -\,0.1 & +\,0.1 & -\,0.1 & +\,0.1 & -\,0.1 & +\,0.1 \\ 
		\hline 	
		$\varepsilon_{e\mu}$ & 4418 & 4425 & 2001 & 2017 & 4420 & 4440 & 2002 & 2019 \\		
		\hline		
		$\varepsilon_{e\tau}$ & 4449 & 4404 & 2019 & 2005 & 4448 & 4422 & 2020 & 2006 \\ 	
		\hline
		$\varepsilon_{\mu\tau}$ & 4492 & 4427 & 2015 & 2038 & 4514 & 4443 & 2019 & 2040 \\ 	
		\hline \hline
	\end{tabular}
\mycaption{The total number of expected reconstructed $\mu^-$ and $\mu^+$ events using 500 kt$\cdot$yr exposure of the ICAL detector over 10 years for both the three-layered (w/ core) and two-layered (w/o core) density profiles of Earth with NC-NSI parameters, $\varepsilon_{e\mu}$, $\varepsilon_{e\tau}$, and $\varepsilon_{\mu\tau}$. For the standard three-layered profile in the absence of NSI, the total number of reconstructed $\mu^-$ ($\mu^+$) events are 4418 (2012). Similarly, for the standard two-layered profile, the total number of reconstructed events are 4427 (2013) with SI. We use the three-flavor neutrino oscillation parameters as given in table~\ref{tab:osc-param-value}, assuming NO.}
\label{tab:events_vc}
\end{table}

To quantify the statistical significance for validating a core inside Earth in the presence of NSI, we scale down the reconstructed events from a 1000-year MC simulation to a 10-year MC simulation, corresponding to a 500 kt$\cdot$yr exposure of ICAL. The expected event rates are estimated using the reconstructed muons in the energy ($E_\mu^\text{rec}$) range of 1 to 25 GeV, zenith angle ($\cos\theta_\mu^\text{rec}$) range of $-\,1$ to 1, and the reconstructed hadrons in the energy (${E^\prime}_{\text{had}}^\text{rec}$) range of 0 to 25 GeV. The total number of expected reconstructed $\mu^-$ ($\mu^+$) events under SI scenario would be approximately 4418 (2012) with a 500 kt$\cdot$yr exposure of ICAL. This estimation considers the three-flavor neutrino oscillations in the presence of Earth's matter effects, assuming the standard three-layered density profiles of Earth and normal mass ordering. We consider the benchmark values of neutrino oscillation parameters as given in table~\ref{tab:osc-param-value}. In table~\ref{tab:events_vc}, we present the expected $\mu^-$ and $\mu^+$ events using 500 kt$\cdot$yr exposure of the ICAL detector in the presence of flavor-violating NC-NSI parameters, $\varepsilon_{e\mu}$, $\varepsilon_{e\tau}$, and $\varepsilon_{\mu\tau}$, for three-layered and two-layered density profiles of Earth assuming NO. We observe that the total event rates for Earth density profiles with and without a core are not significantly different for all three NSI parameters. However, the binned event distributions for both the Earth density profiles are expected to differ significantly, which would be reflected in the sensitivity of ICAL to distinguish between these density profiles.

\begin{table}[htb!]
	\centering
	\begin{tabular}{| c | c | c |  c | c | c | c | c | c |} 
		\hline \hline
		& \multicolumn{8}{c|}{Modified $R_\text{CMB}$} \\
		\cline{2-9}
		& \multicolumn{4}{c|}{Smaller core (SC)} & 
		\multicolumn{4}{c|}{Larger core (LC)}  \\
		\cline{2-9}
		& \multicolumn{2}{c|}{$\mu^-$ events} & 
		\multicolumn{2}{c|}{$\mu^+$ events} &  \multicolumn{2}{c|}{$\mu^-$ events} & 
		\multicolumn{2}{c|}{$\mu^+$ events}  \\
		\cline{2-9}
		& -\,0.1 & +\,0.1 & -\,0.1 & +\,0.1 & -\,0.1 & +\,0.1 & -\,0.1 & +\,0.1 \\ 
		\hline 	
		$\varepsilon_{e\mu}$ & 8836 & 8867 & 4009 & 4043 & 8856 & 8870 & 4014 & 4044 \\		
		\hline		
		$\varepsilon_{e\tau}$ & 8909 & 8823 & 4045 & 4014 & 8912 & 8825 & 4047 & 4020 \\ 	
		\hline
		$\varepsilon_{\mu\tau}$ & 9006 & 8844 & 4035 & 4087 & 9018 & 8889 & 4039 & 4088 \\ 	
		\hline \hline
	\end{tabular}
	\mycaption{The total number of expected reconstructed $\mu^-$ and $\mu^+$ events using 1 Mt$\cdot$yr exposure of the ICAL detector over 20 years for SC and LC scenarios in the presence of NC-NSI parameters, $\varepsilon_{e\mu}$, $\varepsilon_{e\tau}$, and $\varepsilon_{\mu\tau}$. For the SC scenario in the absence of NSI, the total number of reconstructed $\mu^-$ ($\mu^+$) events are 8844 (4030). Similarly, for the LC scenario, the total number of reconstructed events is 8862 (4034) with SI. We use the three-flavor neutrino oscillation parameters as given in table~\ref{tab:osc-param-value}, assuming NO.}
	\label{tab:events_cmb}
\end{table}

Similarly, to estimate the sensitivity to locate the CMB in the presence of NSI, we scale down the reconstructed events from a 1000-year MC simulation to a 20-year MC simulation, which is equivalent to a 1 Mt$\cdot$yr exposure of the ICAL detector. The total expected number of observed $\mu^-$ ($\mu^+$) reconstructed events for the standard CMB radius would be approximately 8850 (4032) with 1 Mt$\cdot$yr exposure of ICAL. This estimation considers the three-flavor neutrino oscillations in the presence of standard matter effects with NO, assuming the three-layered density profiles of Earth. We consider the benchmark values of neutrino oscillation parameters as given in table~\ref{tab:osc-param-value}. In table~\ref{tab:events_cmb}, we show the expected $\mu^-$ and $\mu^+$ events using 1 Mt$\cdot$yr exposure of the ICAL detector in the presence of NC-NSI parameters, $\varepsilon_{e\mu}$, $\varepsilon_{e\tau}$, and $\varepsilon_{\mu\tau}$, for SC and LC scenarios assuming NO. Again, we observe that the total event rates for SC and LC scenarios are almost identical for the corresponding NSI parameters. However, after binning these events, the distribution of event difference between the standard $R_\text{CMB}$ and SC or LC scenario would look different and, hence, give rise to the sensitivity of the ICAL detector to locate the CMB with NSI.

\subsection{Differences in the reconstructed event distributions}

In this section, we study the impact of the Earth density profiles with and without a core, as well as a density profile with modified $R_\text{CMB}$ radius, on the bin-wise distribution of reconstructed $\mu^-$ events at the ICAL detector. To illustrate the distribution of event differences, we employ a binning scheme with 10 bins in $E^\text{rec}_\mu$ in the range [1, 25] GeV, which includes 5 bins of 1 GeV each in the range [1, 5] GeV, 1 bin of 2 GeV in the range [5, 7] GeV, 1 bin of 3 GeV in the range [7, 10] GeV, and 3 bins of 5 GeV each in the range [10, 25] GeV. On the other hand, $\cos\theta^\text{rec}_\mu$ consists of 20 uniform bins spanning the range of $-\,1$ to 1. We integrate events over hadron energies (${E^\prime}_{\text{had}}^\text{rec}$) within the range [0, 25] GeV.  Note that this binning scheme differs from what we will use for our analyses, which is described in section~\ref{sec:statistical analysis}. Since the standard matter effects are significant for neutrinos, if the true mass ordering is NO, we decide to show the distribution of event difference only for reconstructed $\mu^-$ events.

\begin{figure}[htb!]
    \centering
    \includegraphics[width=1.0\linewidth]{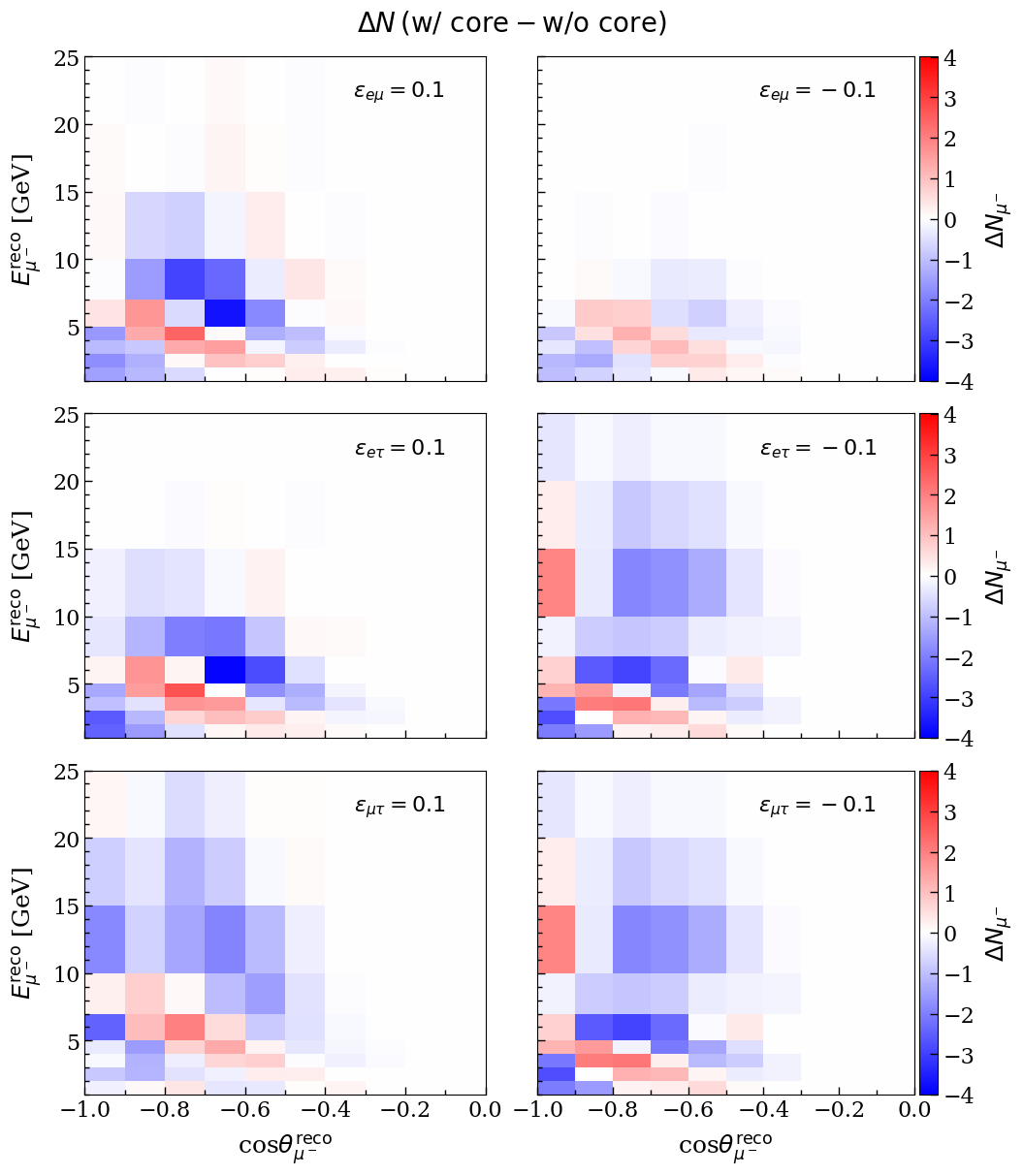}
    \mycaption{Distributions of reconstructed $\mu^-$ event differences between the Earth density profiles with and without a core in the $(E^{\text{reco}}_{\mu^-}, \cos\theta^{\text{reco}}_{\mu^-})$ plane with NSI considering 500 kt$\cdot$yr exposure of the ICAL detector. The top, middle, and bottom rows correspond to the NC-NSI parameters, $\varepsilon_{e\mu}$, $\varepsilon_{e\tau}$, and $\varepsilon_{\mu\tau}$, respectively. The left (right) columns are for the positive (negative) values of these NC-NSI parameters.  We consider the benchmark values of neutrino oscillation parameters as given in table~\ref{tab:osc-param-value}, assuming NO.}
    \label{fig:vc_event_diff_distribution}
\end{figure}

Figure~\ref{fig:vc_event_diff_distribution} shows the event differences between the Earth density profile with and without a core in the plane of $(E^\text{rec}_{\mu^-},\cos\theta^\text{rec}_{\mu^-})$ with NSI for 500 kt$\cdot$yr exposure of ICAL assuming NO. The top, middle, and bottom rows correspond to non-zero values of the NC-NSI parameters $\varepsilon_{e\mu}$, $\varepsilon_{e\tau}$, and $\varepsilon_{\mu\tau}$, respectively. The left (right) columns correspond to the positive (negative) values of these NC-NSI parameters. We can observe that the impact of the flavor-violating NC-NSI parameters on the event difference distributions aligns with their effects on probability differences, as discussed in section~\ref{sec:oscillograms_vc}. The regions of event differences are more pronounced for positive values of $\varepsilon_{e\mu}$ and $\varepsilon_{e\tau}$ compared to their negative counterparts. In contrast, $\varepsilon_{\mu\tau}$ produces similar event difference regions for both positive and negative values. For all three NSI parameters, the event differences are more significant at higher baselines, particularly for the core-passing trajectories.

\begin{figure}[htb!]
	\centering
	\includegraphics[width=1.0\linewidth]{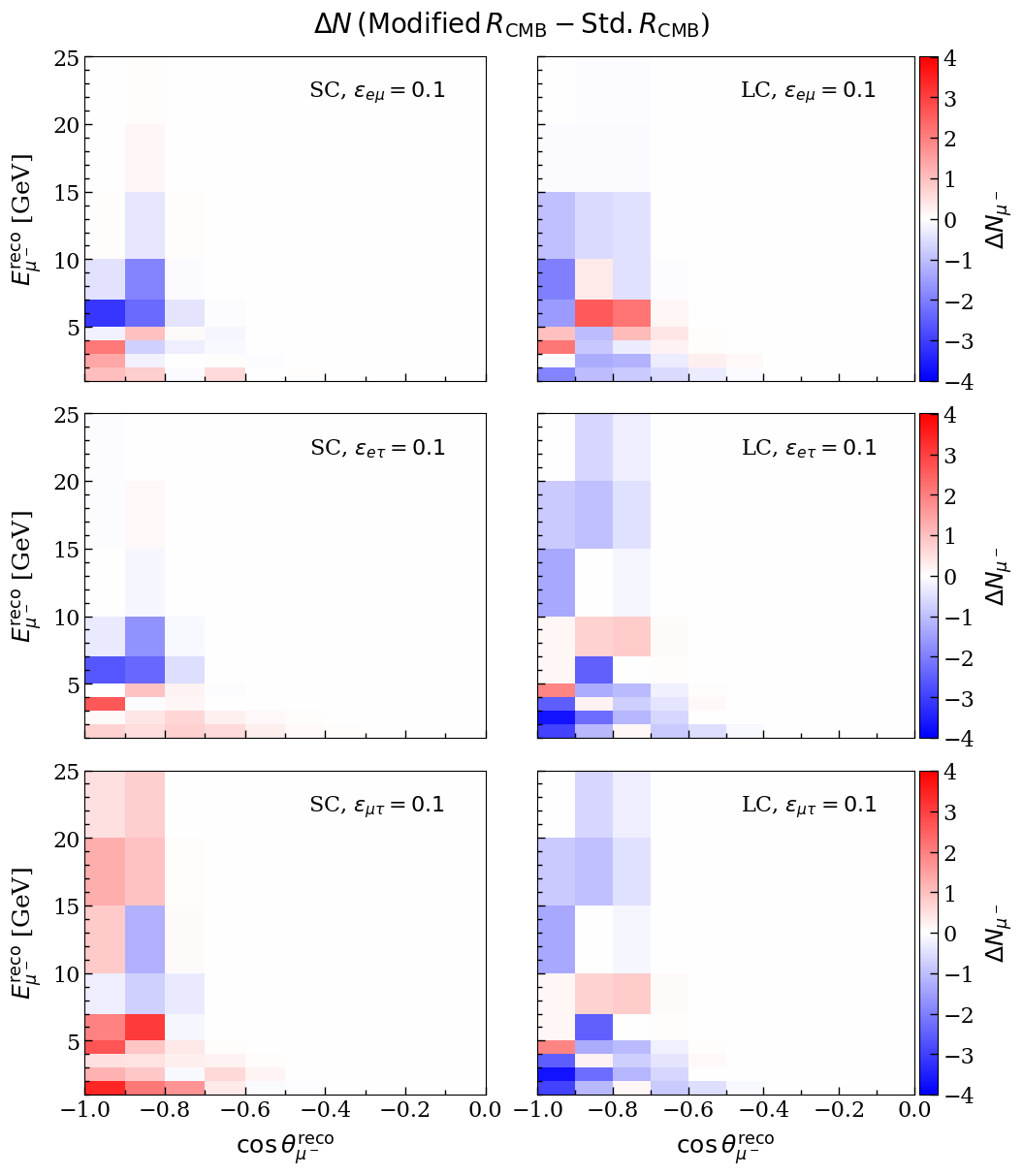}
	\mycaption{Distributions of reconstructed $\mu^-$ event differences between the standard and modified $R_\text{CMB}$ in the  $(E^{\text{reco}}_{\mu^-}, \cos\theta^{\text{reco}}_{\mu^-})$ plane with NSI considering 1 Mt$\cdot$yr exposure of the ICAL detector. The top, middle, and bottom rows correspond to the positive values of NC-NSI parameters, $\varepsilon_{e\mu}$, $\varepsilon_{e\tau}$, and $\varepsilon_{\mu\tau}$, respectively. We consider the benchmark values of neutrino oscillation parameters as given in table~\ref{tab:osc-param-value}, assuming NO.}
	\label{fig:cmb_event_diff_distribution}
\end{figure}

Similarly, the left (right) panels of figure~\ref{fig:cmb_event_diff_distribution} present the distributions of event differences between the standard core and that of the smaller (larger) core in $(E^\text{rec}_{\mu^-},\cos\theta^\text{rec}_{\mu^-})$ plane with NSI for 1 Mt$\cdot$yr exposure of ICAL assuming NO. The top, middle, and bottom rows correspond to the NC-NSI parameters $\varepsilon_{e\mu}$, $\varepsilon_{e\tau}$, and $\varepsilon_{\mu\tau}$, respectively. We present the event difference for only positive values of these NSI parameters. From figure~\ref{fig:cmb_event_diff_distribution}, it is observed that the impacts of the flavor-violating NC-NSI parameters on the event difference distributions align with their effects on the probability differences, as discussed in section~\ref{sec:oscillograms_cmb}. The event differences are primarily concentrated in the core region. However, some differences are also observed in the mantle region due to the angular smearing caused by the discrepancy between the directions of incoming neutrinos and reconstructed muons.

\section{Analysis method}
\label{sec:statistical analysis}

In this section, we discuss the statistical method used to estimate the sensitivity of the ICAL detector for validating a core and measuring its location in the presence of NSI. We calculate Asimov sensitivity~\cite{Cowan:2010js} by performing a $\chi^2$ analysis within the frequentist framework~\cite{Blennow:2013oma}. The MC events are binned using the optimized binning schemes. For the analysis to validate the core, we employ the optimized binning scheme\footnote{For $E^\text{rec}_\mu$, we have a total of 12 bins in which 6 bins of 0.5 GeV in the range ($1-4$) GeV, 3 bins of 1 GeV in the range ($4-7$) GeV, 1 bin of 4 GeV in the range ($7-11$) GeV, and 2 bins of 5 GeV in the range ($11-21$) GeV. For $\cos^\text{rec}_\mu$, we take a total of 21 bins in the range ($-\,1, 1$), where 12 bins of bin width 0.05 in the range ($-\,1.0, -\,0.4$), 4 bins of bin width 0.1 in the range ($-\,0.4, 0.0$), and 5 bins of bin width 0.2 in the range ($0.0, 1.0$). At the same time, we have a total of 4 bins for ${E'}_\text{had}^\text{rec}$ in the range ($0-25$) GeV.} from ref.~\cite{Kumar:2021faw}, whereas for locating the CMB radius, we use the optimized binning scheme\footnote{For $E^\text{rec}_\mu$, we have a total of 16 bins in which 10 bins of 0.5 GeV in the range ($1-6$) GeV, 3 bins of 2 GeV in the range ($6-12$) GeV, 1 bin of 3 GeV in the range ($12-15$) GeV, and 2 bins of 5 GeV in the range ($15-25$) GeV. For $\cos^\text{rec}_\mu$, we take a total of 39 bins in the range ($-\,1, 1$), where 12 bins of bin width 0.0125 in the range ($-\,1.0, -\,0.85$), 18 bins of bin width 0.025 in the range ($-\,0.85, -\,0.4$), 4 bins of bin width 0.1 in the range ($-\,0.4, 0.0$), and 5 bins of bin width 0.2 in the range ($0.0, 1.0$). At the same time, we have a total of 4 bins for ${E'}_\text{had}^\text{rec}$ in the range ($0-25$) GeV.} from ref.~\cite{Upadhyay:2022jfd}. For both analyses, we define the following Poissonian $\chi^2$~\cite{Baker:1983tu} in terms of the reconstructed observables for $\mu^-$ events; $E_\mu^\text{rec}$, $\cos\theta_\mu^\text{rec}$, and ${E^\prime}_{\text{had}}^\text{rec}$, as considered in ref.~\cite{Devi:2014yaa}:
\begin{equation}\label{eq:chisq_mu-}
\chi^2_- = \mathop{\text{min}}_{\xi_l} \sum_{i=1}^{N_{{E'}_\text{had}^\text{rec}}} \sum_{j=1}^{N_{E_{\mu}^\text{rec}}} \sum_{k=1}^{N_{\cos\theta_\mu^\text{rec}}} \left[2(N_{ijk}^\text{theory} - N_{ijk}^\text{data}) -2 N_{ijk}^\text{data} \ln\left(\frac{N_{ijk}^\text{theory} }{N_{ijk}^\text{data}}\right)\right] + \sum_{l = 1}^5 \xi_l^2\,,
\end{equation}
with 
\begin{equation}
N_{ijk}^\text{theory} = N_{ijk}^0\left(1 + \sum_{l=1}^5 \pi^l_{ijk}\xi_l\right)\,.
\label{eq:chisq_2}
\end{equation}
Here, $N_{ijk}^\text{theory}$ and $N_{ijk}^\text{data}$ denote the expected and observed number of reconstructed $\mu^-$ events in a given ($E^{\text{rec}}_\mu$, $\cos\theta^{\text{rec}}_\mu$, $E^{\prime \text{rec}}_{\text{had}}$) bin, respectively. $N_{E_{\mu}^\text{rec}}$, $N_{\cos\theta_\mu^\text{rec}}$, and $N_{{E'}_\text{had}^\text{rec}}$ represent the total number of bins for the reconstructed observables $E_\mu^\text{rec}$, $\cos\theta_\mu^\text{rec}$, and ${E^\prime}_{\text{had}}^\text{rec}$, respectively. The variable $N_{ijk}^0$ indicates the number of expected events in a specific bin without accounting for systematic uncertainties. In this study, we consider the following five systematic uncertainties~\cite{Ghosh:2012px,Thakore:2013xqa}: (i) 20\% uncertainties on flux normalization, (ii) 10\% uncertainty on cross section, (iii) 5\% energy dependent tilt error in flux, (iv) 5\% zenith angle dependent tilt error in flux, and (v) 5\% overall systematics. These systematic uncertainties are incorporated using the well-known method of pulls~\cite{Gonzalez-Garcia:2004pka,Huber:2002mx,Fogli:2002pt}. The variables $\xi_l$ in Eqs.~\ref{eq:chisq_mu-} and \ref{eq:chisq_2} represent the pull parameters associated with the $l^\text{th}$ systematic uncertainty, while $\pi^l_{ijk}$ quantifies the corresponding fractional rate of change in the expected event count with respect to $\xi_l$ i.e., $N^0_{ijk}\pi^l_{ijk} = \left.\frac{\partial N^\text{theory}_{ijk}}{\partial \xi_l}\right|_{\xi_l=0}$.

We define $\chi^2_+$ for reconstructed $\mu^+$ events, which will be estimated separately along with $\chi^2_-$ using the same procedure as described above. The resultant median sensitivity of the ICAL detector is calculated by combining the individual contributions of both $\chi^2_-$ and $\chi^2_+$, which is defined as $\chi^2$:
\begin{equation}
\chi^2 = \chi^2_- + \chi^2_+\,.
\end{equation}
We simulate the MC data using the benchmark values of oscillation parameters provided in table~\ref{tab:osc-param-value} as true parameters. In the fit, we minimize the $\chi^2$ with respect to the pull variables $\xi_l$, accounting for the systematic uncertainties and for relevant oscillation parameters along with the tested NSI parameter. We consider the uncertainties on the atmospheric mixing angle $\sin^2\theta_{23}$ in the range $(0.36 - 0.66)$, the atmospheric mass-squared difference $|\Delta m^2_{\text{eff}}|$ in the range $(2.1 - 2.6)\times 10^{-3}$ eV$^2$, and both choices of neutrino mass orderings, NO and IO. We vary the tested NSI parameter in the range $[-\,0.15, +\,0.15]$. During the minimization, we keep the solar oscillation parameters $\sin^2 2\theta_{12}$ and $\Delta m^2_{21}$ fixed at their true values as given in table~\ref{tab:osc-param-value}. We consider a fixed value for reactor mixing angle $\sin^2 2\theta_{13}$ = 0.0875 both in MC data and theory, as it is well measured~\cite{Capozzi:2025wyn,Esteban:2024eli,NuFIT,Capozzi:2021fjo,deSalas:2020pgw,DayaBay:2022orm}. In the ICAL detector, more than 98\% of the events arise from the disappearance channel $(\nu_{\mu}\rightarrow\nu_{\mu})$. Since the $\delta_\text{CP}$- dependent terms in $P(\nu_{\mu}\rightarrow\nu_{\mu})$ are suppressed by a factor of $(\Delta m^2_{21}/\Delta m^2_{31})\sin\theta_{13}$~\cite{Akhmedov:2004ny}, which is about 0.005, the effect of $\delta_\text{CP}$  on muon neutrino events at ICAL is expected to be small. Therefore, we set $\delta_\text{CP} = 0$ as a benchmark choice in both the MC data and the theory.

\section{Results}
\label{sec:results}
In the first part of this section, we present the sensitivity of the ICAL detector to rule out a two-layered density profile of Earth without the core with respect to a three-layered density profile with the core in the presence of NSI. To evaluate the statistical significance, we simulate prospective MC data with NSI, assuming the three-layered density profile as the true Earth density profile. For the prospective MC data, we consider the benchmark values of neutrino oscillation parameters as given in table~\ref{tab:osc-param-value}. The statistical significance for ruling out the two-layered density profile with respect to the three-layered density profile is defined as:
\begin{equation}
\Delta \chi^2_{\text{core}} = \chi^2 \text{(w/o core, NSI)} - \chi^2 \text{(w/ core, NSI)} \, ,
\label{eq:chi2_vc}  
\end{equation}
where, $\chi^2 \text{(w/o core, NSI)}$ and $\chi^2 \text{(w/ core, NSI)}$ are obtained by fitting the prospective MC data with the two-layered and three-layered density profiles, respectively. Since the three-layered profile represents the true Earth density profile in this analysis, $\chi^2 \text{(w/ core, NSI)} \sim 0$ due to the absence of statistical fluctuations.

In the second part of this section, we present the sensitivity of the ICAL detector to determine the location of the core-mantle boundary. Similar to the previous analysis, we simulate prospective MC data in the presence of NSI, assuming the three-layered density profile with the standard CMB location as the true Earth density profile. The statistical significance for measuring the location of the CMB is defined as:
\begin{equation}
\Delta \chi^2_{\text{CMB}} = \chi^2 \text{(modified}\, R_\text{CMB}\text{, NSI)} - \chi^2 \text{(standard}\, R_\text{CMB}\text{, NSI)} \, ,
\label{eq:chi2_cmb}   
\end{equation}
where, $\chi^2 \text{(modified}\, R_\text{CMB}\text{, NSI)}$ and $\chi^2 \text{(standard}\, R_\text{CMB}\text{, NSI)}$ are estimated by fitting the prospective MC data with the density profiles having modified and standard $R_\text{CMB}$ in theories, respectively. Since the true Earth density profile corresponds to the standard CMB location, $\chi^2 \text{(standard}\, R_\text{CMB}\text{, NSI)}$ $\sim 0$ when statistical fluctuations are suppressed.

\subsection{Sensitivity to validate Earth's core in the presence of NSI}
\label{sec:results_vc}

\begin{figure}[htb!]
	\centering
	\includegraphics[width=1.0\linewidth]{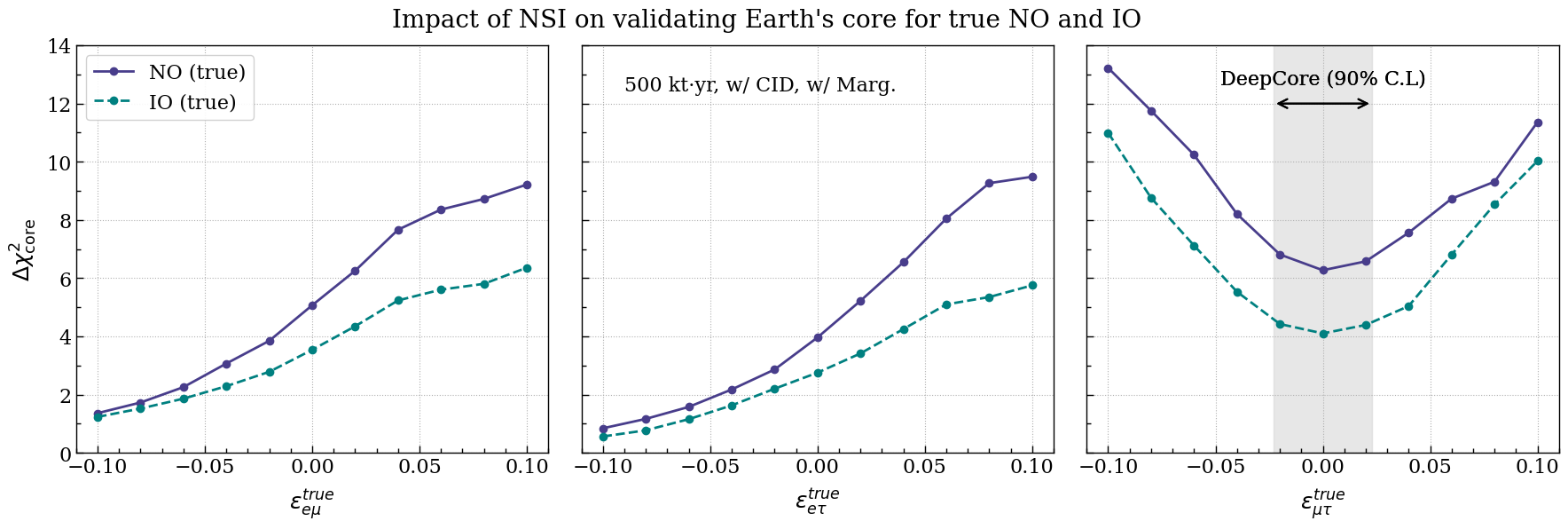}
	\mycaption{The $\Delta \chi^2_\text{core}$ for validating the core inside Earth by rejecting the two-layered coreless density profile as a function of true values of NC-NSI parameters taken one at a time. The left, middle, and right columns correspond to the NC-NSI parameters $\varepsilon_{e\mu}$, $\varepsilon_{e\tau}$, and $\varepsilon_{\mu\tau}$, respectively. The solid (dashed) curve represents the sensitivity calculated, assuming the true mass ordering to be NO (IO). For each plot, we minimize over $\sin^2\theta_{23}$, $\Delta m^2_\text{eff}$, both mass orderings, and corresponding NSI parameter in the fit. We also minimized over all the systematic uncertainty parameters (see section~\ref{sec:statistical analysis}). We consider 500 kt$\cdot$yr exposure of the ICAL detector with CID capability. For the prospective MC data, we consider the benchmark values of neutrino oscillation parameters as given in table~\ref{tab:osc-param-value}. The light gray band in the right panel demonstrates the current bound on $\varepsilon_{\mu\tau}$ at 90\% C.L. from the IceCube DeepCore experiment~\cite{IceCubeCollaboration:2021euf}.}
	\label{fig:result_vc_mass_ordering}
\end{figure}

The sensitivity to validate Earth's core is quantified in the presence of the flavor-violating NC-NSI parameters $\varepsilon_{e\mu}$, $\varepsilon_{e\tau}$, and $\varepsilon_{\mu\tau}$ considered one at a time where we vary only one of these NSI parameters and keep others fixed at zero. In figure~\ref{fig:result_vc_mass_ordering}, the $\Delta \chi^2_{\text{core}}$ present the Asimov sensitivity for ruling out the two-layered coreless profile with respect to the three-layered profile as a function of the true values of the NC-NSI parameters $\varepsilon^\text{true}_{e\mu}$ (left panel), $\varepsilon^\text{true}_{e\tau}$ (middle panel), and $\varepsilon^\text{true}_{\mu\tau}$ (right panel). These sensitivities are calculated by generating the prospective MC data with a given NSI parameter taken in the range $[-\,0.1, +\,0.1]$, while we minimize over $\sin^2\theta_{23}$, $\Delta m^2_\text{eff}$, both mass orderings, and the corresponding NSI parameter in theory. We also minimized over all the systematic uncertainty parameters (see section~\ref{sec:statistical analysis}). The $\Delta \chi^2_{\text{core}}$ value at $\varepsilon^\text{true}_{\alpha\beta}=0$ represents the SI case. The sensitivities are calculated by considering 500 kt$\cdot$yr exposure of the ICAL detector with CID capability. The light gray band in the right panel demonstrates the current bound on $\varepsilon_{\mu\tau}$ at 90\% C.L. from the IceCube DeepCore experiment~\cite{IceCubeCollaboration:2021euf}. Since the present bound on $\varepsilon_{e\mu}$ and $\varepsilon_{e\tau}$ span the complete range of $-\,0.1$ to $+\,0.1$, we have not shown the gray bands corresponding to these NSI parameters.

In figure~\ref{fig:result_vc_mass_ordering}, we can observe that the sensitivity to rule out the two-layered coreless density profile with respect to the three-layered density profile increases with the true values of the NC-NSI parameter $\varepsilon_{e\mu}$ (or $\varepsilon_{e\tau}$) while varying the true value from negative to positive. This linear effect of these NSI parameters can be explained by eq.~\ref{eq:Pmue-mat-NSI} in appendix~\ref{app:Peu-NSI}. On the other hand, the sensitivity for the case of $\varepsilon_{\mu\tau}$ is symmetric around the SI case or $\varepsilon^\text{true}_{\mu\tau} = 0$. This effect of $\varepsilon_{\mu\tau}$ on the expected sensitivity can be explained by eq.~\ref{eq:Pmuu-mat-NSI} in appendix~\ref{app:Puu-NSI}. In conclusion, the presence of NSI would significantly impact the sensitivity for rejecting a coreless density profile of Earth.

\begin{figure}[htb!]
	\centering
	\includegraphics[width=1.0\linewidth]{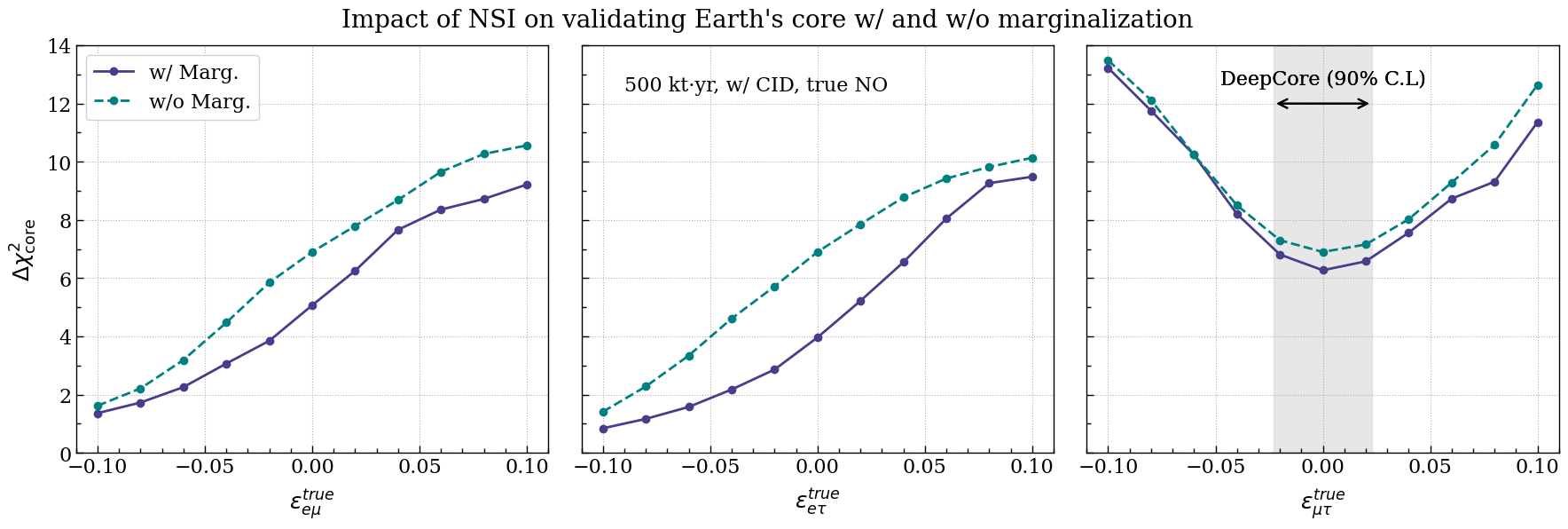}
	\mycaption{The $\Delta \chi^2_\text{core}$ for validating a core inside Earth by rejecting the two-layered coreless density profile as a function of true values of NC-NSI parameters taken one at a time. The left, middle, and right columns correspond to the NC-NSI parameters $\varepsilon_{e\mu}$, $\varepsilon_{e\tau}$, and $\varepsilon_{\mu\tau}$, respectively. Each panel shows the comparison of the sensitivity with minimization over the oscillation and the corresponding NSI parameter (solid curve) and without minimization over these parameters (dashed curve). In both the cases, we minimized over all the systematic uncertainty parameters (see section~\ref{sec:statistical analysis}). We consider 500 kt$\cdot$yr exposure of the ICAL detector with CID capability. For the prospective MC data, we consider the benchmark values of neutrino oscillation parameters as given in table~\ref{tab:osc-param-value} with NO. The light gray band in the right panel demonstrates the current bound on $\varepsilon_{\mu\tau}$ at 90\% C.L. from the IceCube DeepCore experiment~\cite{IceCubeCollaboration:2021euf}. }
	\label{fig:result_vc_marg}
\end{figure}

Figure~\ref{fig:result_vc_mass_ordering} also shows the impact of the true neutrino mass ordering on the Asimov sensitivities, where solid (dashed) curves correspond to the true mass ordering to be NO (IO). We can observe that the sensitivity for IO is smaller than that for NO, which can be explained by the lower statistics of antineutrino events as compared to neutrino events because the interaction cross section for antineutrinos is approximately three times smaller than that of neutrinos. Figure~\ref{fig:result_vc_marg} shows the impact of the uncertainties of the above-mentioned oscillation and the corresponding  NSI parameter on the Asimov sensitivities, where solid (dashed) curves correspond to the sensitivities with (without) minimization. Note that the minimization over systematic parameters is still performed for both curves. The minimization over oscillation and the corresponding NSI parameter in the fit reduces the sensitivity for validating the core for the case of $\varepsilon_{e\mu}$ and $\varepsilon_{e\tau}$. On the other hand, the impact of minimization on the sensitivity is minor for the case of $\varepsilon_{\mu\tau}$.

\subsection{Sensitivity to locate the CMB in the presence of NSI}
\label{sec:results_cmb}

\begin{figure}[htb!]
	\centering
	\includegraphics[width=1.0\linewidth]{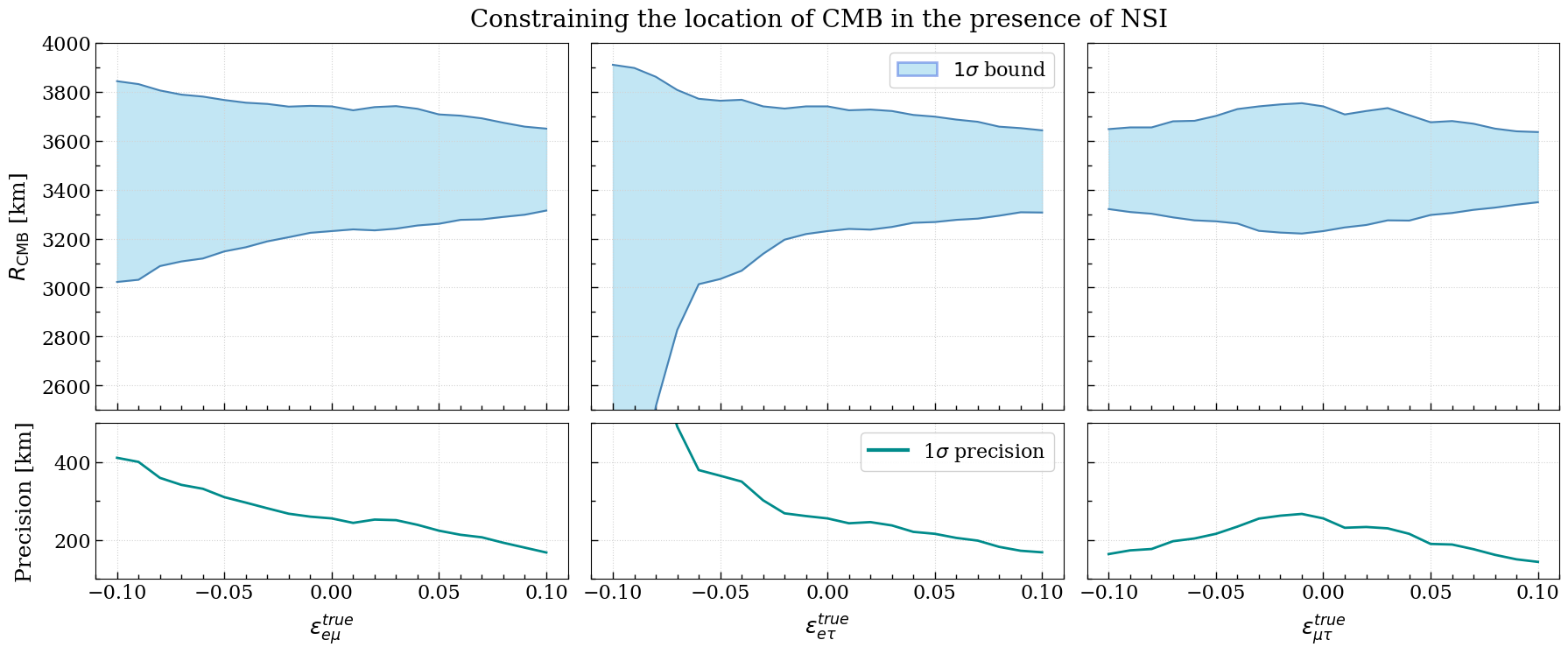}
	\mycaption{The $1\sigma$ allowed ranges of the CMB location (shaded bands) as functions of the true values of the NC-NSI parameters $\varepsilon_{e\mu}$, $\varepsilon_{e\tau}$, and $\varepsilon_{\mu\tau}$ considering one at a time. The solid curves in the bottom panels represent the  $1\sigma$ precision on the CMB location. The left, middle, and right columns correspond to the true NSI parameters $\varepsilon_{e\mu}$, $\varepsilon_{e\tau}$, and $\varepsilon_{\mu\tau}$, respectively. The oscillation and NSI parameters are kept fixed at their true values in the fit. However, we minimized over all the systematic uncertainty parameters (see section~\ref{sec:statistical analysis}). We consider 1 Mt$\cdot$yr exposure of the ICAL detector with CID capability. For the prospective MC data, we consider the benchmark values of neutrino oscillation parameters as given in table~\ref{tab:osc-param-value} with NO.}
	\label{fig:result_measuring_cmb}
\end{figure}

In this section, we calculate the Asimov sensitivity to measure the position of CMB radius $R_\text{CMB}$ in the presence of the flavor-violating NC-NSI parameters $\varepsilon_{e\mu}$, $\varepsilon_{e\tau}$, and $\varepsilon_{\mu\tau}$ considering one at a time. Figure~\ref{fig:result_measuring_cmb} presents the $1\sigma$ bounds (top panels) and $1\sigma$ precisions (bottom panels) on the CMB location as functions of the true values of NC-NSI parameters $\varepsilon_{e\mu}$ (left panels), $\varepsilon_{e\tau}$ (middle panels), and $\varepsilon_{\mu\tau}$ (right panels). The sensitivities in each panel are calculated by considering 1 Mt$\cdot$yr exposure of the ICAL detector with CID capability, assuming the true neutrino mass ordering to be NO. Since the minimization of the oscillation and NSI parameters does not show any impact on this sensitivity, we have kept these parameters fixed at their true values in the fit. However, we minimized over all the systematic uncertainty parameters (see section~\ref{sec:statistical analysis}). The bands in the top panels of figure~\ref{fig:result_measuring_cmb} show the $1\sigma$ bounds on the $R_\text{CMB}$ radius, while the curves in the bottom panels demonstrate the $1\sigma$ precisions on $R_\text{CMB}$ radius. The method to obtain these $1\sigma$ bounds from sensitivity curves for a representative choice of $\varepsilon^\text{true}_{\alpha\beta}=\pm \, 0.1$ is described in appendix~\ref{app:cmb_results}.

The $1\sigma$ bound or the precision on $R_\text{CMB}$ radius improves (deteriorates) compared to the SI case for the positive (negative) values of $\varepsilon_{e\mu}$ and $\varepsilon_{e\tau}$. On the other hand, the $1\sigma$ bound or the precision on $R_\text{CMB}$ radius always improves for $\varepsilon_{\mu\tau}$ irrespective of the sign of this NSI parameter. Again these effects of $\varepsilon_{e\mu}$ and $\varepsilon_{e\tau}$ on sensitivity can be explained by eq.~\ref{eq:Pmue-mat-NSI} in appendix~\ref{app:Peu-NSI}, and effect of $\varepsilon_{\mu\tau}$ can be explained by eq.~\ref{eq:Pmuu-mat-NSI} in appendix~\ref{app:Puu-NSI}. In conclusion, the presence of NSI would impact the measurement of the location of the core-mantle boundary.

\section{Summary and concluding remarks}
\label{sec:conclusion}

As we are in the era of precise measurements of neutrino oscillation parameters, it has become feasible to utilize neutrino oscillations to explore the inner structure of Earth. The insights obtained through the weak interactions of neutrinos complement those provided by the traditional methods such as seismic studies and gravitational measurements. This approach is broadly referred to as neutrino tomography of Earth. At present, neutrinos seem to be an independent and complementary tool paving the way for the ``multi-messenger tomography of Earth''.

The multi-GeV atmospheric neutrinos traversing through Earth experience matter effects due to the interactions with the ambient electrons, which alter their oscillation patterns. In particular, the mantle-passing neutrinos experience the MSW resonance. At the same time, while passing through the high-density core, neutrinos experience an additional resonance called the PR/NOLR. These PR/NOLR matter effects depend on the density jump at the CMB and its location. A detailed study of the impact of these matter effects on neutrino oscillation probabilities can provide valuable information about the internal structure of Earth. Additionally, the presence of any BSM physics could induce sub-leading effects on neutrino oscillations. These BSM interactions can introduce an additional matter potential, which can further modify the oscillation patterns of Earth-passing neutrinos. In this work, we aim to examine the impact of one such BSM scenario, the NSI, on inferring the presence of a high-density core inside Earth and locating the CMB.

We study the impact of NSI parameters on oscillation probabilities of neutrinos that pass through Earth. We present the differences between the disappearance probabilities  $P(\nu_{\mu} \rightarrow \nu_{\mu})$ using the  Earth density profiles with and without the core in the presence of NSI. A comparison of the probability difference oscillograms between a non-zero NSI parameter case and the standard interaction case reveals that the NSI-induced matter effects can influence neutrino oscillation tomography studies. A similar probability difference oscillogram is observed when comparing modified $R_\text{CMB}$ with the standard $R_\text{CMB}$, further highlighting the impact of NSI.

We perform a statistical analysis to compute the Asimov sensitivity to validate the presence of the core inside Earth. In order to do so, we simulate the prospective MC data using the standard three-layered density profile of Earth with NSI. The statistical significance to reject a two-layered coreless Earth density profile with respect to the three-layered density profile in the presence of NSI is calculated in terms of median $\Delta \chi^2$ by fitting the prospective MC data. We quantify $\Delta \chi^2$ for a range of true values of a given NSI parameter. The computed $\Delta \chi^2$ values differ from those in the SI case, and the trends as functions of NSI parameters are consistent with their influence on neutrino oscillation patterns. We observe that if true mass ordering is NO, the ICAL would be able to validate the presence of the core inside Earth with $\Delta \chi^2$ of 6.3 for a SI scenario. Furthermore, this sensitivity is affected by the presence of NSI. It reaches $\Delta \chi^2$ values of 9.2 (1.3), 9.4 (0.84), and 11.3 (13.2) for NC-NSI parameters $\varepsilon_{e\mu}$, $\varepsilon_{e\tau}$, and $\varepsilon_{\mu\tau}$, respectively, with their true values to 0.1 ($-\,0.1$) considering them one at a time. 

To locate $R_\text{CMB}$ radius, the statistical significance is quantified by generating prospective MC data based on the standard three-layered density profile with the standard $R_\text{CMB}$ radius in the presence of NSI and fitting it with the density profile having modified $R_\text{CMB}$. We observe that the NSI impacts the precision of the CMB location. For the NC-NSI parameters, $\varepsilon_{e\mu}$, $\varepsilon_{e\tau}$, and $\varepsilon_{\mu\tau}$, the precision improves for positive values. However, for $\varepsilon_{\mu\tau}$, the precision is slightly better than the SI case, regardless of the sign. At the $1\sigma$ confidence level, the precision is $\pm \, 165$ km ($\pm \, 410$ km), $\pm \, 165$ km (no bound), and $\pm \, 160$ km ($\pm \, 140$ km) for $\varepsilon_{e\mu}$, $\varepsilon_{e\tau}$, and $\varepsilon_{\mu\tau}$, respectively, with NSI parameter values of 0.1 ($-\,0.1$) considering one at a time. However, for the SI scenario, the precision is $\pm \, 250$ km at the 1$\sigma$ confidence level.

Ongoing and upcoming atmospheric neutrino experiments, such as ORCA, Hyper-Kamiokande, DUNE, IceCube/DeepCore/Upgrade, and P-ONE, with their high-statistics neutrino data, are expected to place strong constraints on NSI parameters. Precise knowledge of both NSI and oscillation parameters is crucial for neutrino oscillation tomography studies. Our analysis demonstrates that the presence of NSI significantly impacts both the validation of the Earth's core and the determination of the core-mantle boundary location.

\subsubsection*{Acknowledgements}
We would like to thank A. Dighe and S. Goswami for their useful comments and constructive suggestions on our work. S.K.A. and J.K. acknowledge the financial support from the Swarnajayanti Fellowship (sanction order no. DST/SJF/PSA- 05/2019-20) provided by the Department of Science and Technology (DST), Govt. of India, and the Research Grant (sanction order no. SB/SJF/2020-21/21) provided by the Science and Engineering Research Board (SERB), Govt. of India, under the Swarnajayanti Fellowship project. A.K.U. acknowledges financial support from the DST, Govt. of India (sanction order no. DST/INSPIRE Fellowship/2019/IF190755). The numerical simulations are performed using the ``SAMKHYA: High-Performance Computing Facility'' at the Institute of Physics, Bhubaneswar, India.

\begin{appendix}

\section{Effect of $\varepsilon_{\mu\tau}$ on $P(\nu_\mu \rightarrow \nu_\mu)$ disappearance channel }
\label{app:Puu-NSI}
The flavor-violating NC-NSI parameter $\varepsilon_{\mu\tau}$ significantly impacts the $P(\nu_\mu \rightarrow \nu_\mu)$ disappearance probability, contributing to the leading order. In contrast, the NC-NSI parameters corresponding to the electron sector, $\varepsilon_{e\mu}$ and $\varepsilon_{e\tau}$, play a sub-dominant role in this channel. The authors in eq.~[35] of ref.~\cite{Kopp:2008} provides an approximate analytical expression for $P(\nu_\mu \rightarrow \nu_\mu)$ that accounts for the NSI parameters $\varepsilon_{\mu\mu}$, $\varepsilon_{\mu\tau}$, and $\varepsilon_{\tau\tau}$ at leading order. We consider this equation to get an approximate expression for $P(\nu_\mu \rightarrow \nu_\mu)$ disappearance channel only in the presence of NC-NSI parameter $\varepsilon_{\mu\tau}$ which occurs during neutrino propagation. We assume $\delta_{\text{CP}}=0$ and focus on real $\varepsilon_{\mu\tau}$  by taking NSI phase $\phi_{\mu\tau} = 0$ or $\pi$, which leads to,
\begin{equation}
\varepsilon_{\mu\tau} \equiv |\varepsilon_{\mu\tau}|\cos(\phi_{\mu\tau}) \, .
\end{equation}
The resulting $P(\nu_\mu \rightarrow \nu_\mu)$ disappearance probability is given by,
{ \footnotesize
	\begin{align}
	P(\nu_\mu \to \nu_\mu) &\simeq 1 - s_{2\times23}^2 \sin^2 \frac{\ldm L}{4E} \nonumber\\
	& - \varepsilon_{\mu\tau} s_{2\times23} \left[ s_{2\times23}^2 \frac{a_{cc} L}{2E} \sin \frac{\ldm L}{2E} + 4 c_{2\times23}^2 \frac{a_{cc}}{\ldm} \sin^2\frac{\ldm L}{4E} \right] \nonumber\\
	& + \mathcal{O}\Big(\frac{\sdm}{\ldm} \Big)
	+ \mathcal{O} (s_{13})
	+ \mathcal{O} ( \eps^2 ) \,,
	\label{eq:Pmuu-mat-NSI}
	\end{align}
}
where, $s_{2\times ij} = \sin 2\theta_{ij}$,
$c_{2\times ij} = \cos 2\theta_{ij}$.

The second term in eq.~\ref{eq:Pmuu-mat-NSI} is driven by the standard interactions. The third term contains the effect of NC-NSI parameter $\varepsilon_{\mu\tau}$ on the $P(\nu_\mu \rightarrow \nu_\mu)$ disappearance probability. We can have the following observations from the third term:
\begin{itemize}
	\item Two terms are proportional to $\varepsilon_{\mu\tau}$. One where the oscillating function is $\sin \frac{\ldm L}{2E}$, while the other oscillating with $\sin^2\frac{\ldm L}{4E}$. Therefore, the first term can be positive or negative depending upon the value of the phase or the neutrino mass ordering. In contrast, the second term is always positive.
	\item The case with NO and $\varepsilon_{\mu\tau} > 0$ is equivalent to the case with IO and $\varepsilon_{\mu\tau} < 0$. Similarly, the case with NO and $\varepsilon_{\mu\tau} < 0$ is equivalent to the case with IO and $\varepsilon_{\mu\tau} > 0$. 
	\item For neutrinos (antineutrinos), the $P(\nu_\mu \rightarrow \nu_\mu)$ survival probability decreases (increases) in the first scenario, while in the second scenario, it increases (decreases).
\end{itemize}

\section{Effect of $\varepsilon_{e\mu}$ and $\varepsilon_{e\tau}$ on $P(\nu_e \rightarrow \nu_\mu)$ appearance channel }
\label{app:Peu-NSI}
The flavor-violating NC-NSI parameters $\varepsilon_{e\mu}$ and $\varepsilon_{e\tau}$ affects the $P(\nu_e \rightarrow \nu_\mu)$ appearance probability channel significantly and contributing to the leading orders. In eq.~[33] of ref.~\cite{Kopp:2008}, authors have given the approximate analytic expression for $P(\nu_\mu \rightarrow \nu_e)$ appearance probability channel in presence of NSI parameters $\varepsilon_{e\mu}$ and $\varepsilon_{e\tau}$. We use this equation to obtain an approximate expression for $P(\nu_e \rightarrow \nu_\mu)$ appearance channel in presence of NC-NSI parameters $\varepsilon_{e\mu}$ and $\varepsilon_{e\tau}$ which appear during the neutrino propagation. To obtain the expression, we assume $\delta_{\text{CP}}=0$ and focus on real $\varepsilon_{\alpha\beta}$  by taking NSI phase $\phi_{\alpha\beta} = 0$ or $\pi$, which implies,
\begin{align}
	\begin{split}
		\varepsilon_{e\mu} &\equiv |\varepsilon_{e\mu}|\cos(\phi_{e\mu}) \\
		\varepsilon_{e\tau} &\equiv |\varepsilon_{e\tau}|\cos(\phi_{e\tau}) \, .
	\end{split}
\end{align}
The resulting expression for $P(\nu_e \rightarrow \nu_\mu)$ appearance channel in presence of NC-NSI parameters $\varepsilon_{e\mu}$ and $\varepsilon_{e\tau}$ is given as:
{ \footnotesize
	\begin{align}
		P(\nu_e \rightarrow \nu_\mu) &\simeq
		4 \tilde{s}_{13}^{2} s_{23}^{2} \sin^{2} \frac{(\ldm - a_{\rm CC})L}{4E}
		\nonumber\\
		&\hspace{0.5cm}
		+ \Big( \frac{\sdm}{\ldm} \Big)^2 c_{23}^2 s_{2 \times 12}^2
		\Big( \frac{\ldm}{a_{\rm CC}} \Big)^2 \sin^{2} \frac{a_{\rm CC} L}{4E}
		\nonumber\\
		&\hspace{0.5cm}
		- \frac{\sdm}{\ldm} \tilde{s}_{13} s_{2 \times 12} s_{2 \times 23}
		\frac{\ldm}{a_{\rm CC}}
		\left[   \sin^{2} \frac{a_{\rm CC} L}{4E}
		- \sin^{2} \frac{\ldm L}{4E}
		+ \sin^{2} \frac{(\ldm - a_{\rm CC})L}{4E}
		\right]                                           \nonumber\\
		&\hspace{0.5 cm}
		- 4 \epsilon_{e\mu} \tilde{s}_{13} s_{23} c_{23}^{2}
		\left[   \sin^{2} \frac{a_{\rm CC} L}{4E}
		- \sin^{2} \frac{\ldm L}{4E}
		+ \sin^{2} \frac{(\ldm - a_{\rm CC})L}{4E}
		\right]                                           \nonumber\\
		&\hspace{0.5 cm}
		+ 8 \epsilon_{e\mu} \tilde{s}_{13} s_{23}^{3} \frac{a_{\rm CC}}{\ldm - a_{\rm CC}}
		\sin^{2} \frac{(\ldm - a_{\rm CC})L}{4E} \nonumber \\
		&\hspace{0.5 cm}
		+ 4 \epsilon_{e\tau} \tilde{s}_{13} s_{23}^{2} c_{23}
		\left[   \sin^{2} \frac{a_{\rm CC} L}{4E}
		- \sin^{2} \frac{\ldm L}{4E}
		+ \sin^{2} \frac{(\ldm - a_{\rm CC})L}{4E}
		\right]                                           \nonumber\\
		&\hspace{0.5 cm}
		+ 8 \epsilon_{e\tau} \tilde{s}_{13} s_{23}^{2} c_{23} \frac{a_{\rm CC}}{\ldm - a_{\rm CC}}
		\sin^{2} \frac{(\ldm - a_{\rm CC})L}{4E}  \nonumber\\
		&\hspace{0.5 cm}
		+ 4 \epsilon_{e\mu} \frac{\sdm}{\ldm} s_{2 \times 12} c_{23}^3 \frac{\ldm}{a_{\rm CC}}
		\sin^{2} \frac{a_{\rm CC} L}{4E}                 \nonumber\\
		&\hspace{0.5 cm}
		- 2 \epsilon_{e\mu} \frac{\sdm}{\ldm} s_{2 \times 12} s_{23}^2 c_{23} \frac{\ldm}{\ldm - a_{\rm CC}}
		\left[   \sin^{2} \frac{a_{\rm CC} L}{4E}
		- \sin^{2} \frac{\ldm L}{4E}
		+ \sin^{2} \frac{(\ldm - a_{\rm CC}) L}{4E}
		\right]                                           \nonumber\\
		&\hspace{0.5 cm}
		- 4 \epsilon_{e\tau} \frac{\sdm}{\ldm} s_{2 \times 12} s_{23} c_{23}^2 \frac{\ldm}{a_{\rm CC}}
		\sin^{2} \frac{a_{\rm CC} L}{4E}                 \nonumber\\
		&\hspace{0.5 cm}
		- 2 \epsilon_{e\tau} \frac{\sdm}{\ldm} s_{2 \times 12} s_{23} c_{23}^2 \frac{\ldm}{\ldm -a_{\rm CC}}
		\left[   \sin^{2} \frac{a_{\rm CC} L}{4E}
		- \sin^{2} \frac{\ldm L}{4E}
		+ \sin^{2} \frac{(\ldm - a_{\rm CC}) L}{4E}
		\right]                                           \nonumber\\
		&\hspace{0.5 cm}
		+ \mathcal{O}\Big( \Big[ \frac{\sdm}{\ldm} \Big]^3 \Big)
		+ \mathcal{O}\Big( \Big[ \frac{\sdm}{\ldm} \Big]^2 s_{13} \Big)
		+ \mathcal{O}\Big( \frac{\sdm}{\ldm} s_{13}^2 \Big)
		+ \mathcal{O} ( s_{13}^3 )                        \nonumber\\
		&\hspace{0.5 cm}
		+ \mathcal{O}\Big( \eps \Big[ \frac{\sdm}{\ldm} \Big]^2 \Big)
		+ \mathcal{O}\Big( \eps s_{13} \frac{\sdm}{\ldm} \Big)
		+ \mathcal{O} ( \eps s_{13}^2 )
		+ \mathcal{O} ( \eps^2 ) \, ,
		\label{eq:Pmue-mat-NSI}
	\end{align}
}
where, $s_{ij} =  \sin\theta_{ij}$,
$c_{ij} = \cos\theta_{ij}$, $s_{2\times ij} = \sin 2\theta_{ij}$,
$c_{2\times ij} = \cos 2\theta_{ij}$. The effective 1-3 mixing angle $\theta_{13}$ in matter is defined as, 
\begin{align}
	\tilde{s}_{13} \equiv \frac{\ldm}{\ldm - a_{\rm CC}} s_{13} + \mathcal{O}(s_{13}^2) \, .
\end{align}

The first three terms in eq.~\ref{eq:Pmue-mat-NSI} are governed by the standard matter potential, with the first term having the dominant effect. The remaining terms incorporate the effects of the NC-NSI parameters $\varepsilon_{e\mu}$ and $\varepsilon_{e\tau}$ on the $P(\nu_e \rightarrow \nu_\mu)$ appearance probability. Notably, the effect of $\varepsilon_{e\mu}$ is primarily contributed by the fifth term,
\begin{equation}
	+ 8 \epsilon_{e\mu} \tilde{s}_{13} s_{23}^{3} \frac{a_{\rm CC}}{\ldm - a_{\rm CC}}\sin^{2} \frac{(\ldm - a_{\rm CC})L}{4E} \,,
\end{equation}
where, the factor ($\ldm - a_{\rm CC}$) in the denominator induces the standard matter-driven resonance effects for neutrinos (antineutrinos) if the true neutrino mass ordering is normal (inverted). This term can be positive (negative) because $a_{\rm CC}$ is positive (negative) for neutrinos (antineutrinos). Consequently,  a positive (negative) value of $\varepsilon_{e\mu}$ increases (decreases) $P(\nu_e \rightarrow \nu_\mu)$ for neutrinos. An equivalent effect is also observed for $\varepsilon_{e\tau}$, which can be understood through the seventh
term in eq.~\ref{eq:Pmue-mat-NSI}.

\section{$P(\nu_\mu \rightarrow \nu_\mu)$ survival probability oscillograms in the presence of NSI}
\label{app:Puu_oscillograms}

\begin{figure}[htb!]
	\centering
	\includegraphics[width=1.0\linewidth]{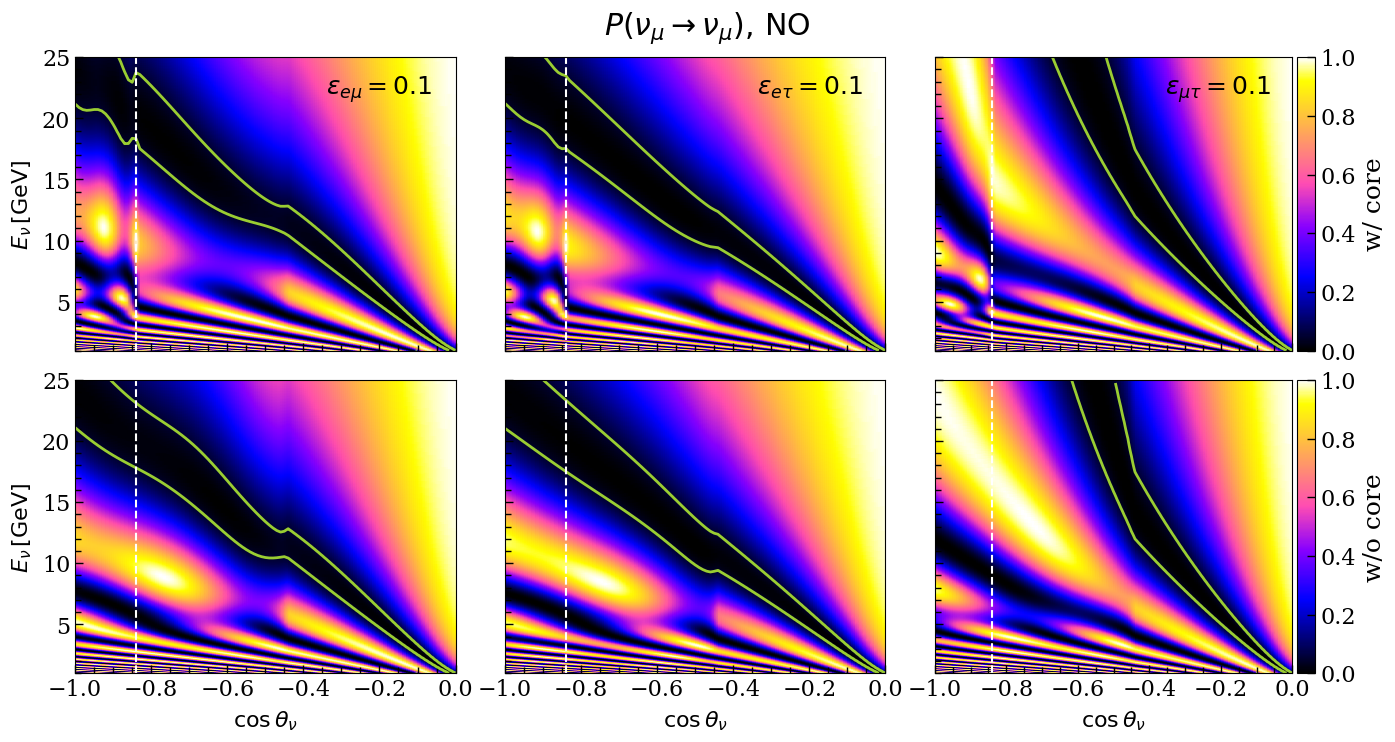}
	\mycaption{The three-flavor $P(\nu_\mu \rightarrow \nu_\mu)$ survival probability oscillograms in $(E_\nu, \cos\theta_\nu)$ plane for Earth density profile with core (top panels) and without core (bottom panels) in the presence of flavor-violating NC-NSI parameters $\varepsilon_{e\mu}$, $\varepsilon_{e\tau}$, and $\varepsilon_{\mu\tau}$ considered one at a time. The vertical dashed-white lines represent the standard CMB location.  The green bands represent the first oscillation valley. We consider the benchmark values of neutrino oscillation parameters as given in table~\ref{tab:osc-param-value} assuming NO.}
	\label{fig:app-vc-mumu}
\end{figure}

In figure~\ref{fig:sm-oscillogram} of section~\ref{sec:NSI_effect}, we present the three-flavor $P(\nu_\mu \rightarrow \nu_\mu)$ survival probability oscillograms in the presence of the SI using three-layered density profile of Earth. In this appendix, we demonstrate the effects of flavor-violating NC-NSI parameters $\varepsilon_{e\mu}$, $\varepsilon_{e\tau}$, and $\varepsilon_{\mu\tau}$ on three-flavor $P(\nu_\mu \rightarrow \nu_\mu)$ survival probabilities, calculated using considered altering Earth density profiles.

In figure~\ref{fig:app-vc-mumu}, we show the three-flavor $P(\nu_\mu \rightarrow \nu_\mu)$ survival probabilities in the plane of $(E_\nu, \cos\theta_\nu)$ with NC-NSI parameters. The left, middle, and right columns correspond to the NC-NSI parameters $\varepsilon_{e\mu}$, $\varepsilon_{e\tau}$, and $\varepsilon_{\mu\tau}$, respectively, with a true value of $0.1$ considered one at a time. We compare the effects of these flavor-violating NC-NSI parameters for the three-layered Earth density profile with a core (top panels) and a coreless two-layered density profile (bottom panels). Compared to the left panel of figure~\ref{fig:sm-oscillogram}, in the top panels of figure~\ref{fig:app-vc-mumu}, we can observe the bending or distortion in the oscillation valley (as represented by the green band) as well as noticeable changes in the MSW and PR/NOLR resonance regions depending upon the NC-NSI parameter considered. A monotonic bending is observed for the NC-NSI parameter $\varepsilon_{\mu\tau}$, which occurs due to an additional term introduced by the matter potential arising from $\varepsilon_{\mu\tau}$. On the other hand, noticeable changes occur in the matter effects region, along with the distortions in the oscillation valley, due to the flavor-violating NC-NSI parameters $\varepsilon_{e\mu}$ and $\varepsilon_{e\tau}$.

In comparison to the top panels of figure~\ref{fig:app-vc-mumu}, if we look at the bottom panels, the PR/NOLR resonance region is not visible, and the MSW resonance region is modified significantly. This indicates that the absence of the core significantly modifies the matter effects, and the PR/NOLR resonance occurs primarily due to the sharp density transition around the CMB ($\cos\theta_\nu = -\,0.84$). However, apart from these standard matter effects changes, other features due to the NC-NSI parameters are still present. In figure~\ref{fig:vc-mutau} and figure~\ref{fig:vc-emu-etau} of section~\ref{sec:oscillograms_vc}, we present the probability differences oscillograms between the top and the bottom panels of figure~\ref{fig:app-vc-mumu}.
 
\begin{figure}[htb!]
	\centering
	\includegraphics[width=1.0\linewidth]{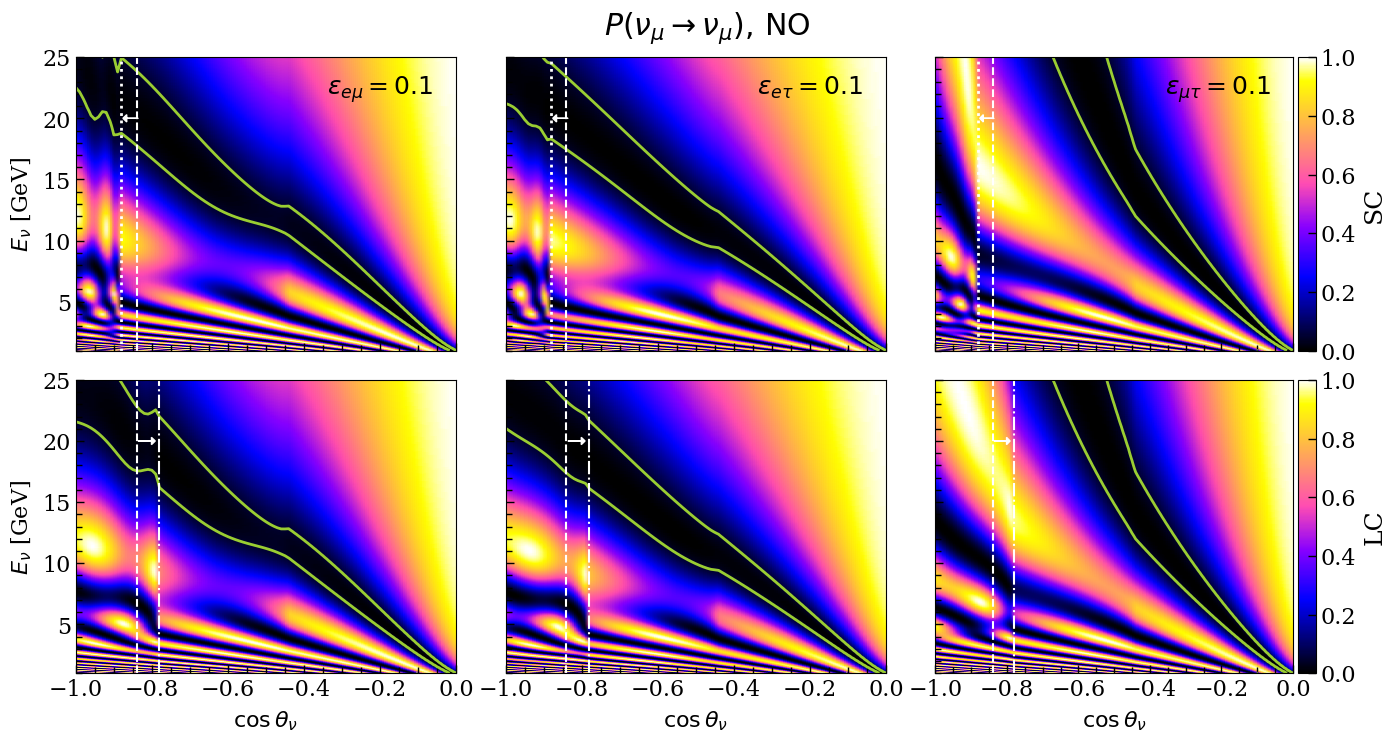}
	\mycaption{The three-flavor $P(\nu_\mu \rightarrow \nu_\mu)$ survival probability oscillograms in $(E_\nu, \cos\theta_\nu)$ plane for Earth density profile with smaller core (top panels) and larger core (bottom panels) in the presence of flavor-violating NC-NSI parameters $\varepsilon_{e\mu}$, $\varepsilon_{e\tau}$, and $\varepsilon_{\mu\tau}$ considered one at a time. The dashed, dotted, and dot-dashed white vertical lines in each panel represent the standard CMB, SC, and LC. The white arrows demonstrate the direction of the CMB modification from its standard location.  The green bands represent the first oscillation valley. We consider the benchmark values of neutrino oscillation parameters as given in table~\ref{tab:osc-param-value} assuming NO.}
	\label{fig:app-cmb-numu}
\end{figure}

In figure~\ref{fig:app-cmb-numu}, we demonstrate the three-flavor $P(\nu_\mu \rightarrow \nu_\mu)$ survival probabilities in the plane of $(E_\nu, \cos\theta_\nu)$ using Earth density profile with smaller core (top panels) and larger core (bottom panels) in the presence of NC-NSI parameters. The left, middle, and right columns correspond to the NC-NSI parameters $\varepsilon_{e\mu}$, $\varepsilon_{e\tau}$, and $\varepsilon_{\mu\tau}$, respectively, with a true value of $0.1$ considered one at a time. To obtain these plots, we modify the $R_\text{CMB}$ radius by $-\,500$ km for SC and $+\,500$ km for LC from its standard value of 3480 km. The features due to the NC-NSI parameters are similar to figure~\ref{fig:app-vc-mumu} except for changes in the MSW and PR/NOLR resonances region due to modification in the position of CMB. Compared to the top panels of figure~\ref{fig:app-vc-mumu}, where the CMB is located at its standard position, the significant changes in the probabilities occur in the PR/NOLR resonance region for SC as well as LC scenarios in figure~\ref{fig:app-cmb-numu}. In figure~\ref{fig:cmb-mutau} of section~\ref{sec:oscillograms_cmb}, we present the probability difference oscillograms between the top (bottom) panels of figure~\ref{fig:app-cmb-numu} and the top panel of figure~\ref{fig:app-vc-mumu} for SC (LC) in the presence of NC-NSI parameter $\varepsilon_{\mu\tau}$.

\section{Sensitivity results to constrain the CMB location with NSI}
\label{app:cmb_results}

While showing our sensitivity results to locate the CMB radius in figure~\ref{fig:result_measuring_cmb} of section~\ref{sec:results_cmb}, we present the $1\sigma$ bound on the $R_\text{CMB}$ radius as a function of true values of NSI parameters $\varepsilon_{\alpha\beta}$. In this appendix, we demonstrate the method to obtain these bounds for a representative choice of $\varepsilon^\text{true}_{\alpha\beta} = \pm \, 0.1$ and 0.

\begin{figure}[htb!]
    \centering
    \includegraphics[width=1\linewidth]{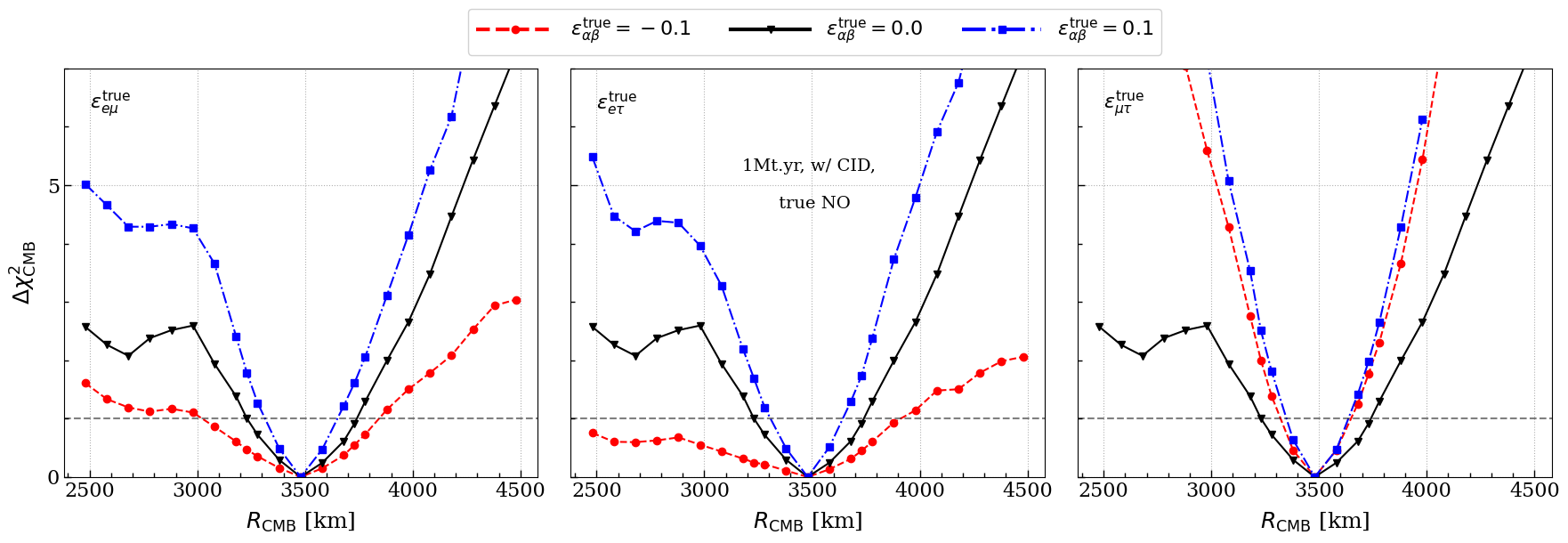}
    \mycaption{The median sensitivity to constrain the CMB location as a function of the $R_\text{CMB}$ radius in the presence of NSI. The left, middle and right panels correspond to $\varepsilon_{e\mu}$, $\varepsilon_{e\tau}$, and $\varepsilon_{\mu\tau}$, respectively, considered one at a time. The dashed red, solid black, and dot-dashed blue sensitivity curves correspond to the NSI parameter true values of $-\,0.1$, $0.0$, and $0.1$, respectively. For all these plots, the oscillation and NSI parameters are kept fixed at their true values in the fit. However, we minimized over all the systematic uncertainty parameters (see section~\ref{sec:statistical analysis}). We consider 1 Mt$\cdot$yr exposure of the ICAL detector with CID capability. For the prospective MC data, we consider the benchmark values of neutrino oscillation parameters as given in table~\ref{tab:osc-param-value} with NO.}
    \label{fig:app_cmb_results}
\end{figure}

Figure~\ref{fig:app_cmb_results} shows the expected sensitivity of the ICAL detector in terms of $\Delta\chi^2_\text{CMB}$ as a function of the position of $R_\text{CMB}$ radius with NSI. The location of $R_\text{CMB}$ radius is modified in theory with respect to the standard $R_\text{CMB}$ radius of 3480 km. The left, middle, and right panels represent the sensitivities calculated for the true values of NC-NSI parameters $\varepsilon_{e\mu}$, $\varepsilon_{e\tau}$, and $\varepsilon_{\mu\tau}$, respectively, considered one at a time. In each panel, solid black, dashed red, and dot-dashed blue sensitivity curves correspond to true values of NC-NSI parameters $\varepsilon^\text{true}_{\alpha\beta} = 0$, $-\,0.1$, and $+\,0.1$, respectively. The horizontal dashed-gray lines represent the Asimov sensitivity at the $1\sigma$ confidence level. These expected sensitivities are quantified for 1 Mt$\cdot$yr exposure of the ICAL detector with its CID capability. We have tested the impact of minimization over the oscillation and NSI parameters and found no effect on the sensitivity; therefore, we have kept them fixed at their true values. However, we minimized over all the systematic uncertainty parameters (see section~\ref{sec:statistical analysis}). This figure shows that the ICAL detector would be able to constrain the CMB location at the $1\sigma$ confidence level with a precision of  $\pm \, 165$ km ($\pm \, 410$ km), $\pm \, 165$ km (no bound), and $\pm \, 160$ km ($\pm \, 140$ km) for $\varepsilon_{e\mu}$, $\varepsilon_{e\tau}$, and $\varepsilon_{\mu\tau}$, respectively, with the true values of $0.1$ ($-\,0.1$). However, for the SI scenario, the precision is $\pm \,250$ km at the 1$\sigma$ confidence level.

\end{appendix}

\bibliographystyle{JHEP}
\bibliography{References}

\end{document}